\documentclass[aps,prx,twocolumn,superscriptaddress,floatfix,nofootinbib]{revtex4-2}
\usepackage[english]{babel} 
\usepackage{amssymb}
\usepackage{amsmath}
\usepackage{mathtools}
\usepackage{txfonts}
\usepackage{mathdots}
\usepackage[normalem]{ulem}
\usepackage[dvips]{graphicx}
\usepackage{epsfig}
\usepackage{graphicx}
\usepackage{array}
\usepackage{amssymb}
\usepackage{amsfonts}
\usepackage{amsmath}
\usepackage{mathrsfs}
\usepackage{booktabs}
\usepackage{threeparttable}
\usepackage{multirow}
\usepackage{subfigure}
\usepackage{epsfig}
\usepackage{threeparttable}
\usepackage{chngpage}
\usepackage{float}
\usepackage{bm}
\usepackage{graphicx}
\usepackage{adjustbox}
\usepackage{comment}
\usepackage{hyperref}
\usepackage{tabularx}
\usepackage{graphicx}
\usepackage{adjustbox}
\usepackage[dvipsnames]{xcolor}

\hypersetup{
    unicode=false,     
    pdftoolbar=false,  
    pdfmenubar=true,   
    pdffitwindow=false, 
    pdfstartview={FitH},
    pdftitle={},    
    pdfauthor={Authors},     
    pdfsubject={},   
    pdfcreator={},   
    pdfproducer={}, 
    pdfkeywords={} {} {}, 
    pdfnewwindow=true,
    colorlinks=true,
    linkcolor=black,
    citecolor=blue, 
    filecolor=magenta,
    urlcolor=blue
}

\newcommand{\beginsupplement}{%
    \setcounter{table}{0}
    \renewcommand{\thetable}{S\arabic{table}}%
    \setcounter{figure}{0}
    \renewcommand{\thefigure}{S\arabic{figure}}%
    \setcounter{equation}{0}
    \renewcommand{\theequation}{S\arabic{equation}}%
    \setcounter{section}{0}
    \renewcommand{\thesection}{S\arabic{section}}%
   }

\makeatletter

\newcommand{\Rmnum}[1]{\expandafter\@slowromancap\romannumeral #1@}
\makeatother

\begin{document}

\title{Origin and emergent features of many-body dynamical localization}

\author{ Ang Yang }
\thanks{These authors contributed equally to this work.}
\affiliation{School of Physics and Zhejiang Key Laboratory of Micro-nano Quantum Chips and Quantum Control, Zhejiang University, Hangzhou 310027, China}

\author{ Zekai Chen}
\thanks{These authors contributed equally to this work.}
\affiliation{Institut f{\"u}r Experimentalphysik und Zentrum f{\"u}r Quantenphysik, Universit{\"a}t Innsbruck, Technikerstra{\ss}e 25, Innsbruck, 6020, Austria}
\affiliation{Department of Physics and Astronomy, University of Rochester, Rochester, New York, 14627, USA}

\author{ Yanliang  Guo }
\email{yanliang.guo@uibk.ac.at}
\affiliation{Key Laboratory of Quantum State Construction and Manipulation (Ministry of Education), School of Physics, Renmin University of China, Beijing 100872, China}
\affiliation{Institut f{\"u}r Experimentalphysik und Zentrum f{\"u}r Quantenphysik, Universit{\"a}t Innsbruck, Technikerstra{\ss}e 25, Innsbruck, 6020, Austria}

\author{ Manuele  Landini}
\affiliation{Institut f{\"u}r Experimentalphysik und Zentrum f{\"u}r Quantenphysik, Universit{\"a}t Innsbruck, Technikerstra{\ss}e 25, Innsbruck, 6020, Austria}

\author{ Hanns-Christoph  N\"agerl}
\affiliation{Institut f{\"u}r Experimentalphysik und Zentrum f{\"u}r Quantenphysik, Universit{\"a}t Innsbruck, Technikerstra{\ss}e 25, Innsbruck, 6020, Austria}

\author{ Lei Ying}
\email{leiying@zju.edu.cn}
\affiliation{School of Physics and Zhejiang Key Laboratory of Micro-nano Quantum Chips and Quantum Control, Zhejiang University, Hangzhou 310027, China}

\begin{abstract}
The question of whether interactions can break dynamical localization in quantum kicked rotor systems has been the subject of a long-standing debate. Here, we introduce an extended mapping from the kicked Lieb-Liniger model to a high-dimensional lattice model and reveal universal features: on-site pseudorandomness and hybrid exponential–algebraic decay couplings with increasing momenta. We find that the exponent and the amplitude of the algebraic decay undergo a crossover as the interaction strength increases. This mapping uncovers the origin of dynamical localization and the interaction effect on the integrability of the system.
An analysis of the generalized fractal dimension and level-spacing ratio supports these findings, highlighting the presence of near integrability and multifractality in different regions of parameter space. Our results offer an explanation for the occurrence of many-body dynamical localization, particularly in strongly correlated quantum gases, and are anticipated to generalize to systems of many particles.
\end{abstract}

\maketitle

\emph{Introduction}---Quantum coherence can give rise to captivating phenomena. For example, the quantum kicked rotor (QKR), as a driven single-particle system, exhibits dynamical localization (DL)~\cite{casati1979quantumpendulum,fishman1982quantumrecurrence,chirikov1986localizationofchaos,Altland1993qkr,Altland1996qftqkr,SANTHANAM2022rotorreview}: nonresonant periodic driving paradoxically leads to suppressed energy absorption. Such an unexpected halt in energy growth in the quantum realm starkly contrasts our everyday experience, which tells us that driven systems generally thermalize to infinite temperature. DL for the QKR can be understood as Anderson localization (AL) in momentum space, characterized by a freezing of the momentum distribution in the course of its evolution~\cite{anderson1958absenceofdiffusion,grempel1984nonintegrable}. This behavior has been studied extensively in cold-atom experiments over the past three decades~\cite{moore1994DL,moore1995rotorexp,raizen1998noiseDL,christensen1998decoherence,delande2008metalinsulatorexp,delande2009metalinsulator,delande2015twodimensionrotor,darcy2001diffusionexp}. However, in reality, particles interact with each other. This raises a fundamental question: Can driven quantum systems avoid thermalization in the presence of interparticle interactions? This challenge has spurred the development and investigation of various many-body QKR models, where particle-particle correlations are introduced through mechanisms such as periodic kicks~\cite{fazio2018CQKRdiffusion,fazio2021CQKRsubdiffusion,victor2017couplerelativistic,victor2016integrableDMBL} or static contact interactions~\cite{konik2020MBDL,vuatelet2021effectivethermal,qin2017interacting}. These models predict a diverse range of behavior, including both DL and delocalized phases. Understanding the mechanism of localization sheds light on the role of many-body interactions in driven quantum systems.

The QKR model with contact interactions has received particular attention since it can be realized in cold-atom platforms. Predictions based on the mean-field approximation suggest that the system delocalizes at long times with a subdiffusive behavior of the energy~\cite{delande2020meanfield,Gligoric2011delocalization,shepelyansky1993nonlinear,shepelyansky2008nonlinearity}, consistent with recent experimental observations~\cite{gupta2022delocalization,david2022delocalization}. However, in one dimension, the mean-field approach runs into severe limitations~\cite{thierry2003onedimension,gogolin2004strongcorrelated}. Beyond this approximation, numerical studies treating two interacting bosons have yielded contradictory results~\cite{qin2017interacting,chicireanu2021fewbodylimit}. In the Tonks-Girardeau (TG) regime, nonperturbative techniques such as the Bose-Fermi mapping~\cite{girardeau1960bosonfermion,buljan2008bosefermimap,rigol2005groundstate,girardeau2005FBmapping,paredes2004tonks,wilson2020observation} predict that DL persists in the interacting many-body situation. This phenomenon has been called many-body dynamical localization (MBDL)~\cite{konik2020MBDL,vuatelet2021effectivethermal,Vuatelet2023dynamicalmanybody}. It has been observed in recent experimental work~\cite{guo2023observation}. The momentum distribution was shown to freeze over the entire range from weak to strong interactions despite the strong drive. 

To date, most studies on MBDL have focused on obtaining the time evolution of the energy and the momentum distribution, starting from low-energy initial states~\cite{konik2020MBDL,vuatelet2021effectivethermal,Vuatelet2023dynamicalmanybody}. However, the microscopic mechanism driving MBDL remains little understood.
In this Letter, we investigate MBDL in the Lieb-Liniger (LL) model~\cite{lieb1963solution1,lieb1963solution2,olshanii1998bosegas,olshanii2001bosegas,rigol2011coldatomreview} augmented by a kicking term~\cite{konik2020MBDL}. Using the LL eigenstates as a basis, we establish an extended mapping from this quantum system to a $N-$dimensional lattice model with hybrid exponential–algebraic decay  couplings, in which the algebraic couplings undergo a crossover as the interaction strength increases. Our analysis reveals the origin of localization and the key factors underlying its potential breakdown. Our conclusion is supported by inspecting the generalized fractal dimensions (GFD) and the energy-level-spacing statistics.

\emph{Model}---We consider the 1D kicked LL model, meant to describe $N$ short-range interacting bosons of mass $m_\mathrm{A}$ constrained to a ring of circumference $L\!=\!2\pi$, subject to a pulsed sinusoidal potential~\cite{konik2020MBDL,chicireanu2021fewbodylimit,vuatelet2021effectivethermal} with wave number $k_{\mathrm{L}}$, as illustrated in Fig.~\ref{schematic}(a). With $k$ numbering the kicks with period $T$, the Hamiltonian is written as the sum of the LL Hamiltonian and the kick Hamiltonian~\cite{konik2020MBDL},
\begin{equation}
    \hat{H}(t) = \hat{H}_{\mathrm{I}} + \sum_{k\in \mathbb{N}}\delta(t-k)\hat{H}_{K},
\end{equation}
\vspace{-0.5cm}
\begin{equation}
    \hat{H}_{\mathrm{I}} = \sum_i^N\frac{\hat{p}_i^2}{2} + g\sum^N_{i<j}\delta\left(\hat{x}_i-\hat{x}_j\right),\quad
     \hat{H}_{\mathrm{K}} = K\sum_i^N \mathrm{cos}\left(\hat{x}_i\right).
\end{equation}
The parameters $g$ and $K$ denote the interaction strength and the kick strength, respectively. For convenience, we represent time in units of $T$, position in units of $1/k_{\mathrm{L}}$, and momentum in units of $\hbar_{\rm{eff}} = 4T\hbar k_{\mathrm{L}}^2/m_\mathrm{A}$, which is an effective Planck constant~\cite{SM}. The various limits are well understood: for $K\!=\!0$, we recover the integrable LL model, which can be solved exactly via the Bethe-Ansatz technique~\cite{lieb1963solution1,lieb1963solution2}. The LL eigenstates $\lvert \mathcal{I}\rangle=\lvert m_1m_2\cdots m_N\rangle$ are labeled by $N$ quasimomenta $\{m_j\}$ with the eigenenergy being $E_{\mathcal{I}}=(\hbar_{\mathrm{eff}}^2/2)\sum_{j=1}^N m_j^2$. The LL parameter $\gamma=gL/(N\hbar^2_{\rm{eff}})$ represents the dimensionless interaction strength. For $K\neq0$, it is known that DL and MBDL occur in the two limits $g\!=\!0$ and $g\!=\!\infty$~\cite{SANTHANAM2022rotorreview,konik2020MBDL}. In both cases the problem can be mapped onto a single-particle problem. 
In the intermediate interaction regime, it is unclear whether the localization phenomenon persists, as illustrated in Fig~\ref{schematic}(b). 

\begin{figure}
\centering
\includegraphics[width=0.95\linewidth]{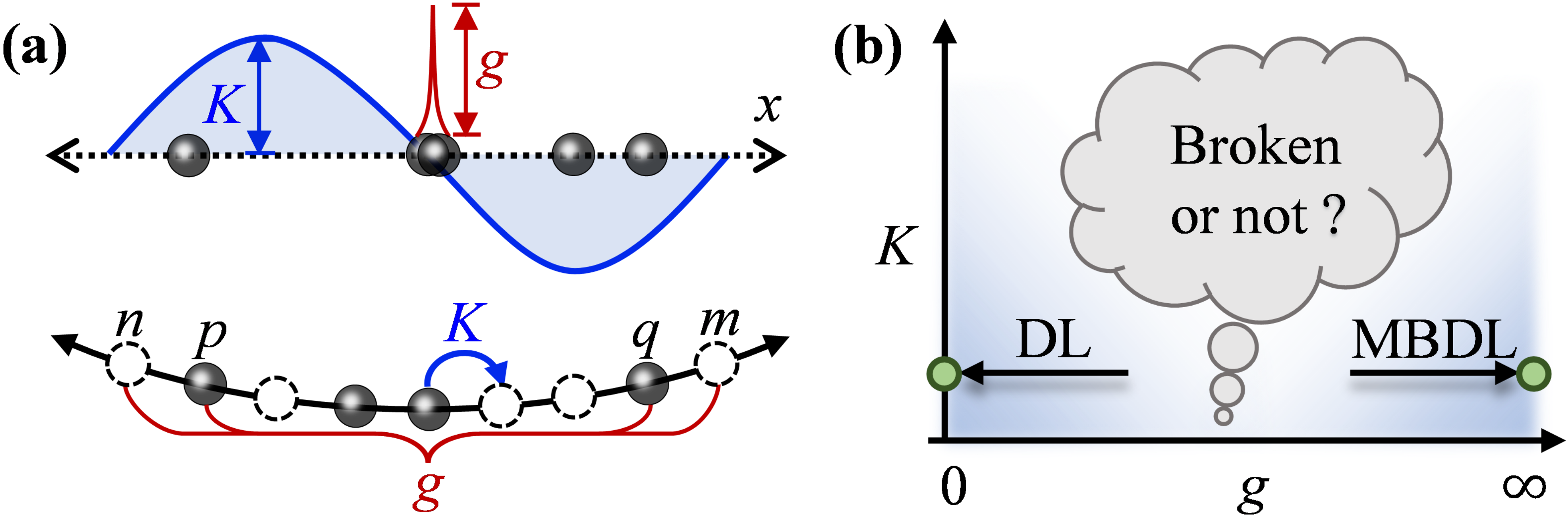}
\caption{
(a) Schematic of the kicked LL system in coordinate (upper) and momentum (lower) space, where blue and red curves denote the pulsed sinusoidal potential and the contact interactions, respectively. In momentum space, the kinetic energy component is represented by a quadratic potential defined over a momentum lattice. The periodic kicks introduce the couplings between adjacent momentum states. Boson-boson interactions result in momentum scattering, where the total momentum is conserved ($n+m=p+q$). (b) Schematic phase diagram of the kicked LL system as a function of the kick strength $K$ and the interaction strength $g$. 
}
\label{schematic}
\end{figure}

The momentum-space representation of the Hamiltonian in second quantization using the Fock basis $\lvert \cdots n_{m}\cdots\rangle$ reads~\cite{SM}
\begin{equation}
    \begin{aligned}
        \hat{H}_{\mathrm{I}} &= \frac{\hbar_\mathrm{eff}^2}{2}\sum_{m=-\infty}^{\infty} {m^2}\hat{b}^\dagger_m\hat{b}_m + \frac{g}{2L}\sum^{\infty}_{m,n,p,q}\hat{b}^\dagger_{m}\hat{b}^\dagger_{n}\hat{b}_{p}\hat{b}_{q}\delta_{m+n,p+q}, \\
        \hat{H}_{\mathrm{K}} &= \frac{K}{2}\sum_{m=-\infty}^{\infty}\left(\hat{b}^\dagger_m\hat{b}_{m+1} + \mathrm{H.c.}\right),
    \end{aligned}
\label{main_hamiltonian}
\end{equation}
where $\hat{b}^\dagger_m$ and $\hat{b}_m$ are the bosonic creation and annihilation operators of momentum state $m$, 
obeying the commutation relations $\left[\hat{b}_m,\hat{b}^\dagger_{m'}\right]=\delta_{m,m'}$. This system satisfies the particle-number conservation condition $\sum_m \hat{b}^{\dagger}_m \hat{b}_m\!=\!N$.  
In the exact diagonalization (ED) approach, if the single-particle momenta are truncated at $|m|\leq M$~\cite{quspin2019,fazio2021CQKRsubdiffusion}, the dimension of the Hilbert space is given by the binomial coefficient $\mathcal{N}=\binom{2M+N}{N}$. We further reduce the Hilbert space by only considering the parity sector of $+1$.

The Floquet operator~\cite{papic2015drivenMBL}, given by $\hat{U} \!= \!\exp\left\{-iT\hat{H}_{\mathrm{I}}/\hbar_\mathrm{eff}\right\}\exp\left\{-i\hat{H}_{\mathrm{K}}/\hbar_\mathrm{eff}\right\}$, describes time evolution in the kicked LL model. Its eigenvalues are complex numbers with unit modulus, giving $\hat{U}=\sum_{\alpha}^{\mathcal{N}}\exp\left\{-i\theta_{\alpha}T/\hbar_\mathrm{eff}\right\}\lvert \phi_{\alpha}\rangle\langle\phi_{\alpha}\rvert$, where the quasienergies $\theta_{\alpha}$ are chosen to be in $\big[-\pi\hbar_\mathrm{eff}/T,\pi\hbar_\mathrm{eff}/T\big)$.

\emph{Mapping to the lattice model}---Next, we outline the mapping from the kicked LL model to a lattice model. By introducing the eigenspectrum decomposition $\hat{H}_{\mathrm{I}}=\sum_{\mathcal{I}}E_{\mathcal{I}}\rvert\mathcal{I}\rangle\langle\mathcal{I}\rvert$ and $\hat{H}_{\mathrm{K}}=\sum_{\mathcal{K}}E_{\mathcal{K}}\rvert\mathcal{K}\rangle\langle\mathcal{K}\rvert$, the eigenvalue equation for the operator $\hat{U}$ can be written as~\cite{SM} 
\begin{align}\label{anderson_eq}
    V_{\mathcal{I}}\Pi_{\mathcal{I}} + \sum_{\mathcal{I}\neq \mathcal{I'}}W_{\mathcal{II'}}\Pi_{\mathcal{I'}} = -W_{{\mathcal{II}}}\Pi_{\mathcal{I}}, 
\end{align}
where
\begin{equation}
    V_{\mathcal{I}}\!=\!\tan\left[\frac{(E_{\mathcal{I}}-\theta_\alpha)T}{2\hbar_{\rm{eff}}}\right],\quad
    W_{{\mathcal{II'}}}\!=\!\sum_{\mathcal{K}} \tan\bigg(\frac{E_{\mathcal{K}}}{2\hbar_{\rm{eff}}}\bigg)\langle \mathcal{I}\lvert\mathcal{K}\rangle\langle\mathcal{K}\rvert\mathcal{I'}\rangle,\nonumber
\end{equation}
and $\Pi_{\mathcal{I}}$ denotes the coefficient containing the information about the Floquet eigenstate. 
The diagonal elements $V_{\mathcal{I}}$ are pseudorandom since $E_{\mathcal{I}}$ has a quadratic dependence on quasimomentum.
The off-diagonal elements $W_{\mathcal{I}\mathcal{I}^\prime}$ represent the couplings between Bethe eigenstates in an $N-$dimensional quasimomentum space spanned by $\{m_j|j=1,2,\dots,N\}$.

\begin{figure}
\centering
\includegraphics[width=1\linewidth]{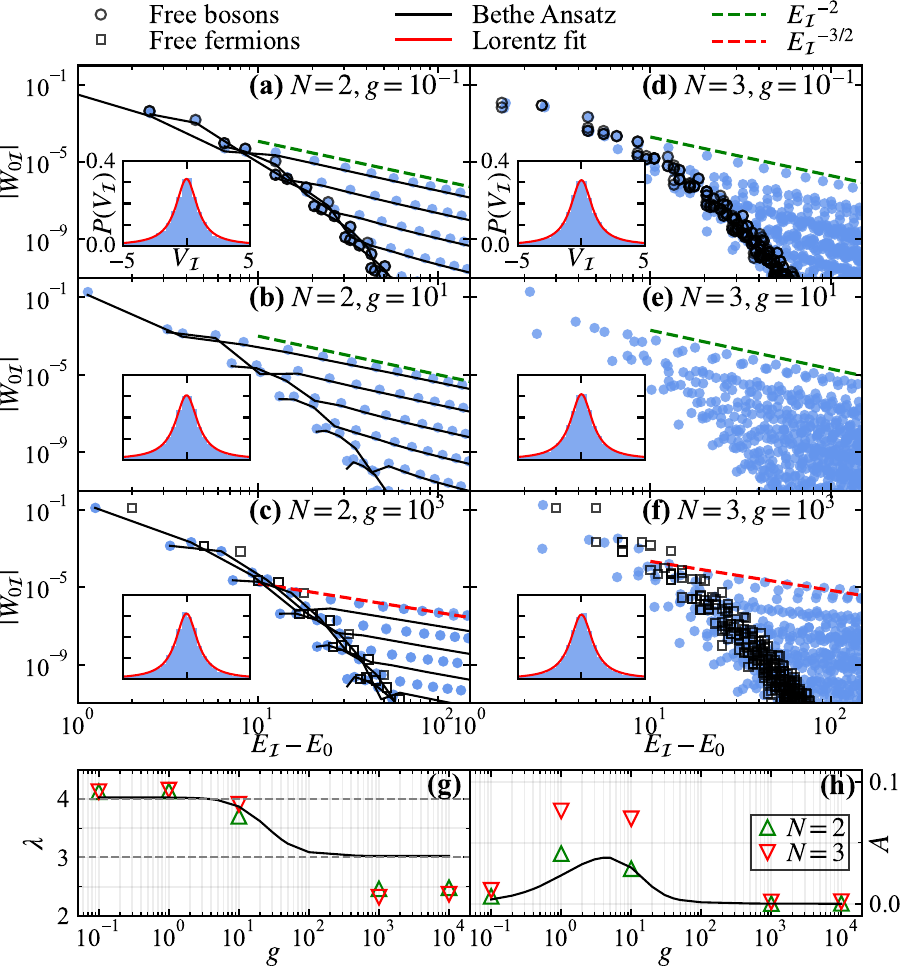}
\caption{
(a-c) The off-diagonal term $\lvert W_{0\mathcal{I}}\rvert$ as a function of energy $E_{\mathcal{I}}-E_0$ for (a) $g=10^{-1}$, (b) $g=10^{1}$, and (c) $g=10^3$ in log-log scale (blue dots). The black lines stand for the Bethe-Ansatz results. The black circles and squares represent the off-diagonal term for the mapped free bosons and fermions, respectively. The green and red dashed lines denote algebraic decay scaling as ${E_{\mathcal{I}}}^{-2}$ and ${E_{\mathcal{I}}}^{-3/2}$, respectively. The insets show the histograms of the on-site term $V_{\mathcal{I}}$, fitted by a Lorentzian distribution. 
The momentum cutoff, the kick strength, and the particle number are $M=100$, $K=0.5$ and $N=2$, respectively. (d-f) show the same data as in (a-c) but for $N=3$ and $M=35$. (g, h) show the decay exponent $\lambda$ and amplitude $A$ as a function of $g$, respectively.
}
\label{mapping_matrix}
\end{figure}

We first employ ED to obtain the LL eigenstates for $N=2,3$.  Starting from the ground state of the LL model, we compute the off-diagonal term $W_{0\mathcal{I}}$ as a function of the energy difference $(E_{\mathcal{I}}-E_0)$ between an excited state $|\mathcal{I}\rangle$ and the ground state $|0\rangle$ for varying interaction strengths, as shown in Fig.~\ref{mapping_matrix}(a-f). Notably, for both $N=2$ and $3$, the off-diagonal terms consistently exhibit hybrid decay: exponential-like decay at low energies and algebraic decay at high energies. Compared to the $N=2$ case, more branches of algebraic decay emerge for $N=3$.
We notice that this exponential-like decay is the one that one would expect for free bosons at $g\rightarrow0$ or for fermions at $g=\infty$. 
The pseudorandom on-site terms consistently follow a Lorentzian distribution (see insets). 
Remarkably, the decay exponent and amplitude of the algebraic tail have a nontrivial dependence on the interaction strength. We further fit the top branch of these algebraic tails of $\lvert W_{0\mathcal{I}}\rvert$ to a power-law function $AE_{\mathcal{I}}^{-\lambda/2}$, and we report the exponent $\lambda$ and the amplitude $A$ for different $g$ in Fig.~\ref{mapping_matrix} (g) and (h). As the interaction strength increases, the decay exponent $\lambda$ shows a crossover from $\lambda=4$ to a lower value around $3$. In parallel, the amplitude $A$ exhibits a nonmonotonic change. This indicates that the variation of the algebraic tails is driven by interactions.

\emph{Structure of the mapped model}---The Bethe-Ansatz technique allows for an analytical description of the mixed decay behavior particularly for $N=2$. The asymptotic behavior of the off-diagonal term is accordingly expressed as~\cite{qin2017interacting,SM}
\begin{equation}
 W_{\mathcal{II'}} \sim \left\{
\begin{aligned}
&\frac{\mathcal{B}_{Q-Q'}^{(2)}(0)}{4\pi^2A_g}\frac{1}{q^4} & & \mathrm{for}\enspace 2qA_g\rightarrow\infty, \,q\gg q', \\
&\frac{\mathcal{B}_{Q-Q'}^{(2)}(0)q'A_g}{\pi^2}\frac{1}{q^3}  & & \mathrm{for}\enspace 2qA_g\rightarrow0, \,\ \ q\gg q',
\end{aligned}
\right.
\label{ansatz_two_particle}
\end{equation}
 where $\mathcal{B}_{Q-Q'}^{(2)}(0)\!=\!\frac{K}{2\hbar_\mathrm{eff}}\int_0^{2\pi}\frac{\cos{(x_\mathrm{c} )}\exp\{2i(Q-Q')x_\mathrm{c}\}}{\cos^2{[K \cos{(x_\mathrm{c})}/\hbar_\mathrm{eff}]}}dx_\mathrm{c}$, $A_g\!=\!\hbar^2_{\mathrm{eff}}/g$, and $x_\mathrm{c}=(x_1+x_2)/2$ is the center-of-mass (c.m.) position.
 $Q\ (q)$ and $Q^\prime\ (q^\prime)$ are the c.m. (relative) momentum of eigenstates $\mathcal{I}$ and $\mathcal{I'}$, respectively. 
 The kicks nontrivially couple to the hopping in the c.m. momentum $\mathcal{B}_{Q-Q'}^{(2)}(0)$, which decays exponentially with $|Q-Q'|$ and indicates DL in the c.m. momentum space.
 Generally, this is expected for any $N$ since $\hat{H}_{\mathrm{I}}$ conserves the c.m. momentum~\cite{SM}. In the relative momentum, there is indeed a crossover in the tail $q^{-\lambda}$ of the hopping $ W_{\mathcal{II'}}$ from $\lambda=4$ to $\lambda=3$ and the tail's amplitude is first proportional, then inversely proportional to $g$ as $g$ grows. 
 As shown in Fig.~\ref{mapping_matrix}, the Bethe-Ansatz results are in agreement with the ED results. Importantly, Eq.~(\ref{ansatz_two_particle}) reveals the mechanism: the kicks lead to exponential decay between LL eigenstates from different c.m. momentum sectors, while interactions further generate an algebraic tail in the relative momentum.

For more particles, particularly for numbers approaching those found in experiments~\cite{guo2023observation}, we approximate $\tan{[({K}/{2\hbar_{\mathrm{eff}}})\sum_j\cos{x_j}]}\approx ({K}/{2\hbar_{\mathrm{eff}}})\sum_j\cos{x_j}$ at small kick strengths. The off-diagonal term becomes~\cite{SM}
\begin{align}\label{W_approx}
    W_{\mathcal{I}\mathcal{I}'}\propto\sum_{h=1}^N\sum_{P',P}A_{P'}^*A_P \sum_{s=\pm 1}\sum_{l=1}^{N+1}\mathop{\mathrm{Res}}_{\omega=-c^{s,PP'}_{h,l}}\left[\frac{e^{-i\omega L}}{\prod_{j=1}^{N+1}(\omega+c^{s,PP'}_{h,j}) }\right], 
\end{align}
where $c^{s,PP'}_{h,j}=\sum_{l=j}^{N}\Delta m^{PP'}_l\!\!+\!s\cdot H(h-j)$ with $\Delta m^{PP'}_l\!=\!m_{P(l)}\!-\!m'_{P'(l)}$ and $H(\cdot)$ being the Heaviside step function. The Bethe coefficient is $A_P\propto (-1)^{\mathrm{sgn}(P)}\prod_{j>l}[m_{P(j)}-m_{P(l)}-ig]$ with $P$ denoting a permutation operation of $N$ quasimomenta $\{m_j\}$.
Here, the kicks effectively couple two different Bethe states whose c.m. momenta satisfy $\sum_{l=1}^N \Delta m^{PP'}_l=2\pi/L$, leading to a second-order pole $c^{s,PP'}_{h,1}=c^{s,PP'}_{h,N+1}$.
For stronger kicks, one only needs to account for additional c.m. momentum sectors regardless of $N$. Moreover, consideration of more particles only introduces additional first-order poles, since $\Delta m^{PP'}_l\neq0$ generally holds for different Bethe states. This analysis suggests that the algebraic tail is universal for arbitrary $N$.

\begin{figure}
\centering
\includegraphics[width=1\linewidth]{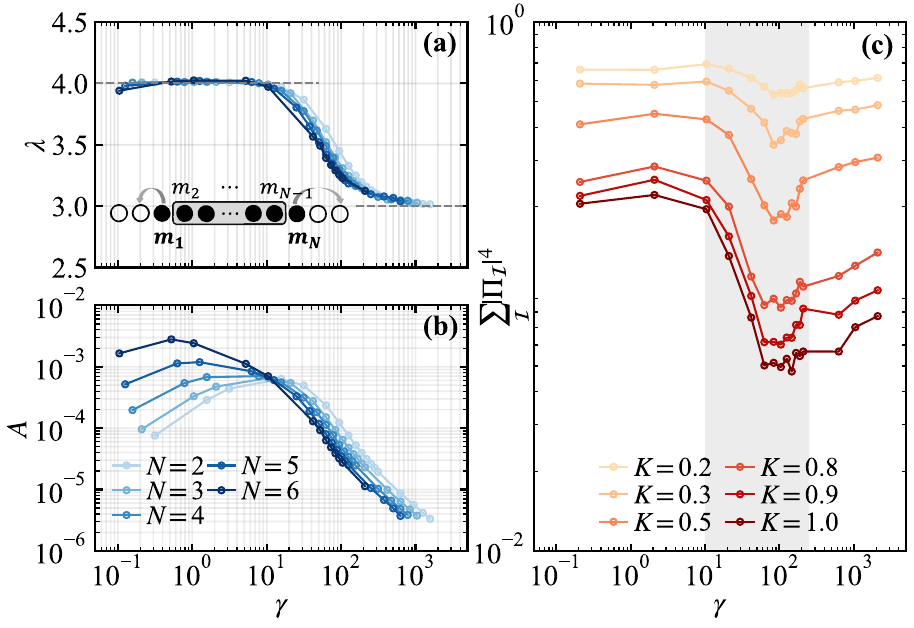}
\caption{
The decay exponent $\lambda$ (a) and the amplitude $A$ (b) as a function of $\gamma$ for different $N$. The inset of (a) illustrates the typical two-particle excitations $\{m_1,m_N\}$ of the excited state $\lvert\mathcal{I}\rangle$, while the other quasimomenta $\{m_2\cdots m_{N-1}\}$ are kept identical to their ground-state value $\lvert0\rangle$. (c) IPR of eigenstates of the mapped model as a function of $\gamma$ for different $K$, which has been averaged over $500$ eigenstates around midspectrum. The momentum cutoff and the particle number are respectively $M=35$ and $N=3$.
}

\label{approx_W}
\end{figure}

For increasing values of $N$, we compute the off-diagonal terms $W_{0\mathcal{I}}$ using Eq.~\eqref{W_approx}, and then fit $W_{0\mathcal{I}}$ to an algebraic function $A{\delta m}^{-\lambda}$~\cite{SM} with $\delta m$ representing the standard deviation of the quasimomentum of the excited state $\lvert\mathcal{I}\rangle$. 
For a typical algebraic tail (see the top branch in Fig.~\ref{mapping_matrix}), only two momenta $m_1$ and $m_N$ vary under the condition $m_1+m_N=2\pi/L$ and $\sum^{N-1}_{j=2}m_j=0$, as illustrated in the inset of Fig.~\ref{approx_W}(a). 
We plot the decay exponent and the amplitude as a function of $\gamma$ in Fig.~\ref{approx_W}. 
The decay exponents for different $N$ converge, exhibiting the identical transition from $4$ to $3$ as seen in Eq.~\eqref{ansatz_two_particle}. The amplitude $A$ changes nonmonotonically with increasing $\gamma$. Generally, for $\lvert \mathcal{I}\rangle$ in each c.m. momentum sector (fixed $\sum^N_{j=1}m_j$), the algebraically decaying branches of $W_{0\mathcal{I}}$ are characteristic of a two-particle excitation (e.g., varying $\{m_1,m_N\}$ with a different set of fixed quasimomenta $\{m_2,\dots,m_{N-1}\}$)~\cite{SM}.
Also, the off-diagonal term exhibits an anisotropy between excited Bethe states $|\mathcal{I}\rangle$ and $|\mathcal{I}^\prime\rangle$~\cite{SM}. To further demonstrate the effect of the algebraic tail on localization, we show the inverse participation ratio (IPR) $\sum_{\mathcal{I}}\lvert\Pi_{\mathcal{I}}\rvert^4$ of the eigenstates of Eq.~\eqref{anderson_eq} in Fig.~\ref{approx_W}(c). All the data consistently exhibit a significant dip (gray shaded) around intermediate $\gamma$, indicating the extended property of eigenstates. Larger kicks greatly decrease the IPR value. Therefore, the mapped model will first transition toward ergodicity at large kicks and intermediate interaction strength.

\emph{Localization and its breakdown---}
The mapping approach uncovers crucial insights about localization. The mapped model forms a complex $N$-dimensional quantum network characterized by a set of degrees of freedom $\{m_1,\dots,m_N\}$. Its diagonal elements exhibit a consistently pseudorandom distribution, constituting the fundamental origin of localization in analogy to the Anderson problem. Interaction induces nontrivial algebraic decay into the off-diagonal elements along specific directions in the high-dimensional network, this famously can thwart localization. Particularly at intermediate interaction strengths, the off-diagonal elements attain maximum amplitude and exhibit the slowest decay, as shown in Fig.~\ref{approx_W}(c). Within this region, large kicks will eventually render the system ergodic. Eq.~\eqref{W_approx} can be equally expressed in terms of the form factor $W_{\mathcal{I}\mathcal{I}'}\propto\langle\mathcal{I}\lvert\Psi^{\dagger}(x)\Psi(x)\rvert\mathcal{I}'\rangle$, where $\Psi(x)$ is the field operator~\cite{SM}. A well-known exact form of the matrix element is available~\cite{korepin1982,Slavnov1989,Caux_2007,Piroli_2015}. Recent work shows that, in the thermodynamic limit, matrix element of the density operator $\Psi^{\dagger}(0)\Psi(0)$ between two Bethe states with extensive excitations decays superexponentially in the system size~\cite{Essler2024prx}. This makes the high-dimensional network effectively disconnected in most directions. We therefore expect that, in the thermodynamic limit, localization will survive, at small $K$ and any value of $g$.

\emph{Statistics of eigenstates}---Here, we apply the traditional diagnostic to the Floquet eigenspectrum and eigenstates to verify the above predictions for the mapped model. We diagonalize $\hat{U}$, focus on the mid spectrum $(\theta_{\alpha}=0)$, typically extracting $500$ eigenstates for various values of the cutoff $M$, ranging from $10$ to $32$. We first study the Floquet eigenstates based on the participation entropies (PE) $S_{\beta}$. Given an eigenstate $\lvert \phi\rangle$ and an $\mathcal{N}$-dimensional basis $\{\lvert\alpha\rangle\}$, the PE are defined as~\cite{nicolas2019MBLfractal,sarkar2023multifractaldimension,lemarie2017fractalscaling,liu2021MBcritical}
$S_{\beta}=\mathrm{ln} \left(\sum_{\alpha=1}^{\mathcal{N}}\lvert\phi_{\alpha}\rvert^{2\beta}\right)/(1-\beta)$ with $\lvert\phi\rangle=\sum_{\alpha=1}^{\mathcal{N}}\phi_{\alpha}\lvert\alpha\rangle$.
Then one can define the GFD $D_{\beta}$ as $D_{\beta}=\lim_{\mathcal{N}\rightarrow\infty}S_{\beta}/\mathrm{ln}\mathcal{N}$. If $D_{\beta}$ nontrivially depends on $\beta$, the state is multifractal, whereas for constant $D_{\beta}$, the state is fractal~\cite{sarkar2023multifractaldimension}. For a fully delocalized state, $D_{\beta}=1$, while $D_{\beta}=0$ represents the scenario of Anderson localizion.  Values of $0<D_{\beta}<1$, indicate an extended but nonergodic state.

\begin{figure}
\centering
\includegraphics[width=1\linewidth]{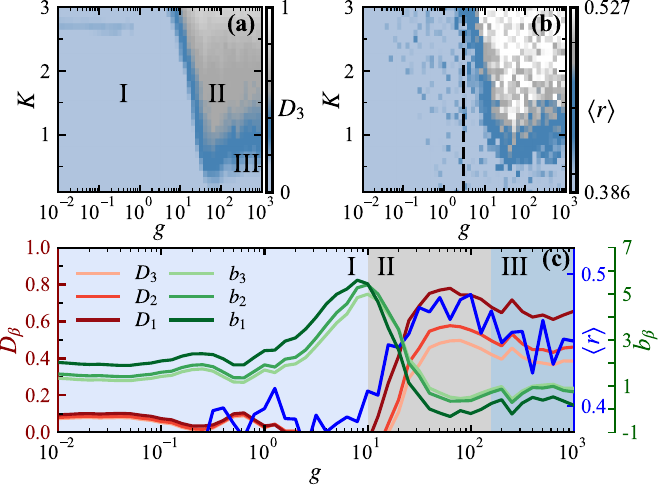}
\caption{
(a) The GFD $D_3$ as a function of $g$ and $K$. (b) The averaged energy-level-spacing ratio $\langle r\rangle$ as a function of $g$ and $K$ for $M=32$. The black dashed line denotes where the LL parameter is $\gamma=1$.
(c) Different $D_{\beta}$, $b_{\beta}$, and $\langle r\rangle$ as a function of $g$ for $K=1.0$. The shaded areas denote different regimes. The number of bosons is $N=3$.
}

\label{phase}
\end{figure}

Fig.~\ref{phase}(a) shows the GFD $D_3$ as a function of $g$ and $K$. It is extracted from a linear fit of the type $\overline{S}_{\beta}=D_{\beta}\mathrm{ln}\mathcal{N}+b_{\beta}$~\cite{SM}. Here, $b_{\beta}$ is a subleading correction at finite dimension. Negative $b_{\beta}$ can be related to a nonergodic volume $\Lambda_{\beta}=\mathrm{exp}\{-b_{\beta}\}$~\cite{nicolas2019MBLfractal}. $\overline{S}_{\beta}$ denotes the averaged $S_{\beta}$ over $500$ eigenstates. One can notice three regimes. We first focus on the small kick strength $K=1$, as shown in Fig.~\ref{phase}(c). When $g$ is small, $D_3\gtrsim0$, the Floquet eigenstate is thus highly localized at regime I. Further increasing $g$ in regime II, $D_3$ exhibits a dramatic increase, while the fitted intercept $b_3$ drops sharply and then levels off near zero. In this regime, the Floquet eigenstates are extended ($0.4\lesssim D_3<1$) but remain nonergodic. As $g$ approaches regime III, $D_3$ decreases. 
This reduction is attributed to the fermionization of the bosons as $g$ approaches infinity, at which point the system behaves like a TG gas~\cite{girardeau2005FBmapping,buljan2008bosefermimap,girardeau1960bosonfermion,rigol2005groundstate}. In this case, the eigenstates are nonergodic across all values of $g$ and the system is in the MBDL phase.
The nonmonotonic change of the GFD aligns with the crossover of the algebraic tail of $W_{\mathcal{I}\mathcal{I}'}$, and is also consistent with the recently predicted nonmonotonic behavior of the interaction energy in the localized regime~\cite{chicireanu2021fewbodylimit}. As we have shown in Fig.~\ref{approx_W}(c), this nonmonotonic change is already present in the mapped model' eigenstates. This indicates a strong link between the structure of the mapped model and the integrability of the kicked LL model. A similar behavior is found for $\beta=1,2$. We also notice that $D_3\neq D_2\neq D_1$, thus the MBDL phase is multifractal. For larger kick strengths, $D_3$ in regime II greatly increases toward the ergodic value, as shown in Fig.~\ref{phase}(a). This suggests the occurrence of delocalization at intermediate interaction strengths, where the algebraic hopping amplitude of the $N-$dimensional lattice reaches a maximum, as we have anticipated.

The same holds for the averaged level-spacing ratio~\cite{haake1991quantumchaos,rigol2014longtimedriven,moessner2014drivingequilibrium} $\langle r\rangle=\mathrm{mean} (r_\alpha)$, where $r_\alpha=\mathrm{min}(\delta_{\alpha+1},\delta_{\alpha})/\mathrm{max}(\delta_{\alpha+1},\delta_{\alpha})$ with $\delta_\alpha=\theta_{\alpha+1}-\theta_\alpha$. As shown in Figs.~\ref{phase}(b) and (c), we find high similarities between the diagrams of $D_3$ and $\langle r\rangle$. 
Regimes I and III correspond to where $\langle r\rangle\approx0.386$, indicating that the system featuring MBDL is nearly integrable in these two regimes. In regime II, however, $\langle r\rangle$ remarkably increases up to around $0.527$, implying the breakdown of localization. We conclude that our results verify the nonmonotonic behavior predicted by the mapped model and establish a phase diagram featuring MBDL and delocalization.

\emph{Discussions}---In summary, we have investigated the microscopic origin of the MBDL phase in the kicked LL model. We find the existence of dynamical localization and its breakdown in different regions of parameter space. Finite interaction strengths introduce an algebraic tail to the localization properties, observed in the mapped model. The exponent and amplitude of this algebraic tail exhibit a nontrivial dependence on the interaction strength, reflecting the characteristic features of the MBDL phase. Further analysis of the many-body eigenstates suggests that MBDL possesses multifractal properties and that finite interactions render the system nonintegrable. Our study provides an explanation for the long-standing question of MBDL's existence at finite interaction strengths and is anticipated to be valid for systems with many particles. 
In End Matter, we propose an experimental implementation to probe the crossover of the algebraic tail. Our results demonstrate that this phenomenon can be observed in the momentum distribution by quenching cold atoms from the noninteracting to the finite-interaction regime.

Since the mapped model corresponds to a high-dimensional problem, a localization transition is expected, as in the 3D case~\cite{anderson1979scaling}. This raises several questions: how does quantum chaos emerge in the kicked LL model at finite interaction strengths? Is there a critical kick strength or interaction strength for an arbitrary number of particles? And is localization stable at finite interaction strength in the thermodynamic limit? 
Despite our explanation for the occurrence of MBDL, future theoretical and experimental investigations are needed. Moreover, the ubiquitous algebraic tail of the momentum distribution, particularly at weak interaction strengths, exhibits a phenomenological resemblance to that observed in quantum turbulence~\cite{ZAKHAROV2001573,ZAKHAROV20041}. This similarity warrants further investigation to substantiate a potential link between these phenomena. 

\emph{Acknowledgments}---We thank Prof. Chushun Tian for many helpful suggestions. We also thank Prof. Marcos Rigol for discussions.  
This work was supported by the National Natural Science Foundation of China (Grants No. 12375021 and No. 12247101), the Zhejiang Provincial Natural Science Foundation of China (Grant No. LD25A050002), the National Key Research and Development Program of China (Grant No. 2022YFA1404203), and the Fundamental Research Funds for the Central Universities (Grant No. lzujbky-2024-jdzx06).
The Innsbruck team acknowledges funding by a Wittgenstein Prize grant with the Austrian Science Fund (FWF) [Grant DOI: 10.55776/Z336] Project No. Z336-N36, by the European Research Council (ERC) with Projects No. 789017 and No. 101201611; by an FFG infrastructure grant with Project No. FO999896041; and by the Austrian Science Fund (FWF)’s [Grant DOI: 10.55776/COE 1] and quantA. Z.C. also acknowledges the support from the University of Rochester.

\emph{Data availability}---The data that support the findings of this article are openly available~\cite{yang_2026_18935863}.

\emph{Note added}---As we were finalizing our manuscript, we found that H. Olsen {\it et al.}~\cite{olsen2025} had independently posted a preprint on the same topic.

\newpage 
\cleardoublepage 
\onecolumngrid
\vspace{+1cm}
\begin{center}
{\Large {\bf End Matter}} 
\end{center}
\vspace{+0.5cm}

\twocolumngrid

\emph{Perspective from the Fock basis analysis}---It is noteworthy that the lattice mapping is done in the LL eigenbasis. To connect to the experimentally accessible momentum distribution, we study the Floquet operator on the Fock basis $\lvert \cdots n_m\cdots\rangle$ and discuss how a crossover of the algebraic tail can be observed experimentally. For arbitrary $N$, we label a family of Fock states by $\lvert\alpha\rangle\equiv\lvert p_\mathrm{c}, \Delta p\rangle$ with their c.m. momentum $p_\mathrm{c}$ and standard deviation $\Delta p$ being
\begin{equation}
    p_\mathrm{c}=\sum_m n_mm,\quad\quad \Delta p={\left(N\sum_m n_mm^2-p_\mathrm{c}^2\right)}^{1/2}.
\end{equation}
The kinetic energy is $E_m\propto p_\mathrm{c}^2+{\Delta p}^2$. We characterize the hopping amplitude in the Fock basis as $\big\lvert U_{p_\mathrm{c}p'_\mathrm{c}}^{\Delta p\Delta p^\prime}\big\rvert=\langle p_\mathrm{c}, \Delta p\lvert\hat{U}\rvert p'_\mathrm{c}, \Delta p'\rangle$. For $N=1$, the hopping amplitude reduces to $\big\lvert U^{\mathrm{S}}_{m0}\big\rvert=(-i)^{m}J_{m}(K/\hbar_{\mathrm{eff}})\exp\{-im^2/2\hbar_{\mathrm{eff}}\}$, where $J_m(\cdots)$ denotes the $m$th-order Bessel function of the first kind. For $N>1$, we consider the hopping amplitude starting from the zero-momentum state $\lvert  0,0\rangle$. There are two cases for increasing kinetic energy: (i) the growth of $p_\mathrm{c}$ while $\Delta p$ remains fixed, leading to the acceleration of all bosons; (ii) the spreading of $\Delta p$ while $p_\mathrm{c}$ remains fixed, facilitating long-range interactions of bosons in momentum space.

The hopping amplitudes of the above two cases are shown in Figure~\ref{floquet_matrix} for three (a) and two (b) particles. As we expected, the hopping amplitude of the c.m. momentum growth $\big\lvert U_{p_\mathrm{c}0}^{00}\big\rvert$ follows a fast decay captured by the noninteracting result (insets), indicating DL. Notably, $\big\lvert U_{p_\mathrm{c}0}^{00}\big\rvert$ is only slightly affected by interactions. The hopping amplitude of $\Delta p$ spreading also exhibits an algebraic tail, falling as $\big\lvert U_{00}^{\Delta p0}\big\rvert\sim {\Delta p}^{-\nu}$. For small $g$, say $g\leq10^1$, the exponent of the tail remains constant as $\nu=2$, and increasing $g$ only affects the overall amplitude. Further increasing the interaction strength up to $g=10^{2}$, $\nu$ dramatically changes to $1$ and stays constant hereafter. The crossover of $\nu$ corresponds to that of $\lambda$. For different fixed $p_\mathrm{c}$, $\big\lvert U_{p_\mathrm{c}p_\mathrm{c}}^{\Delta p0}\big\rvert$ shows a similar behavior~\cite{SM}. A similar crossover of the algebraic tail is found in the occupation spectrum of the one-particle density matrix~\cite{SM}. 

Most importantly, the algebraic tail for $\big\lvert U_{00}^{\Delta p0}\big\rvert$ suggests that kicking a nonequilibrium state such as the zero-momentum state $\lvert0, 0\rangle$ can result in MBDL with different algebraic tails, depending on the interaction strength~\cite{zill2015quenchliebgas}. In the steady-state momentum distribution $n(m)$, this manifests itself as a crossover from $n(m)\propto m^{-4}$ to $m^{-2}$ as the interaction strength increases. 
In most of the previous studies of MBDL, the momentum distribution has consistently exhibited $m^{-4}$ tails when kicking the Lieb-Liniger (LL) ground state. However, the $m^{-2}$ tail associated with kicking a nonequilibrium state has not yet been demonstrated. To verify this, we perform the time evolution, starting from the zero-momentum state for different interaction strengths. Figs.~\ref{floquet_matrix}(c-d) show the localized momentum distribution after $100$ kicks for different interaction strengths and particle numbers at $K=1$. As expected, the momentum distribution $n(m)$ exhibits an algebraic decay as $n(m)\propto m^{-2\nu}$ at large momenta. For $g=1$, the decay exponent is $\nu=2$, reflecting the common signature of MBDL when kicking the LL ground state. A striking deviation from kicking the LL ground state emerges as $\nu$ turns into $1$ at $g=100$. At small $g$, the zero-momentum state closely aligns with the LL ground state, resulting in $n(m) \propto m^{-4}$. As $g \to \infty$, the interaction energy $\langle 0,0 | \hat{H}_{\mathrm{I}} | 0,0 \rangle$ increases significantly, causing the zero-momentum state to deviate substantially from the ground state and equilibrium. Consequently, $n(m) \propto m^{-2}$ when MBDL occurs. Hence we expect that this crossover of the algebraic tail will be experimentally observable in the momentum distribution by quenching the cold-atom system from zero interactions to the finite interaction regime.
\begin{figure}
\centering
\includegraphics[width=1\linewidth]{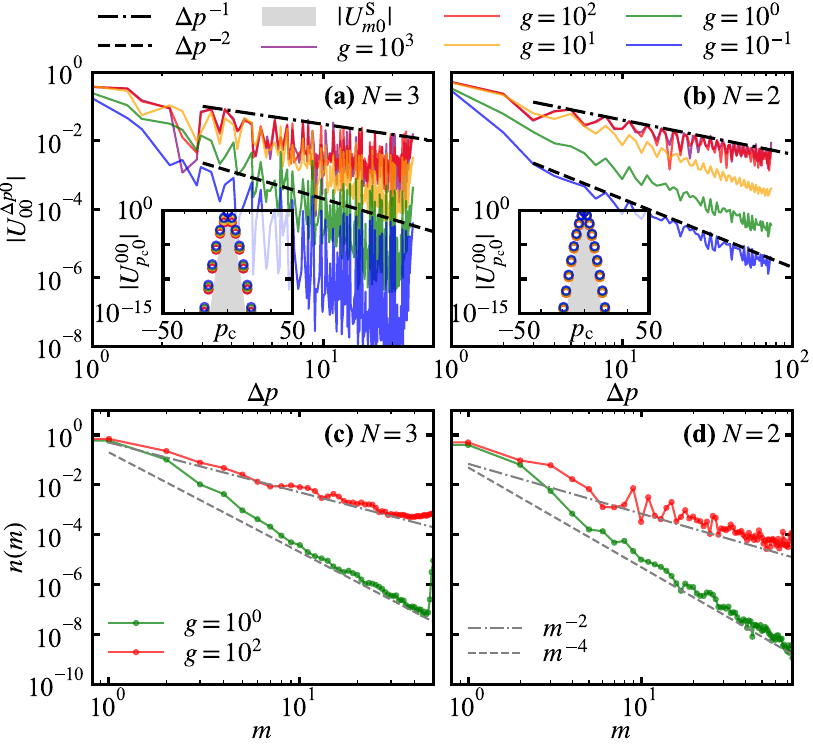}
\caption{
(a) The hopping amplitude as a function of the standard deviation $\Delta p$ for $N=3$ and different values of $g$ in log-log scale. The black dashed line and the black dot-dashed line denote algebraic decay scaling as ${\Delta p}^{-2}$ and ${\Delta p}^{-1}$, respectively. The inset shows the hopping amplitude as a function of the c.m. momentum $p_\mathrm{c}$ in semilog scale. The upper boundary of the gray shaded area denotes the single-particle hopping $\lvert U^{\mathrm{S}}_{m0}\rvert$. The cutoff and the kick strength are $M=30$ and $K=1$, respectively. (b) The same data as in (a) but for $N=2$ and $M=75$. (c) The momentum distribution $n(m)$ after $100$ kicks for the particle number $N=3$ and different interaction strengths $g$ in log-log scale. The cutoff and the kick strength are $M=50$ and $K=1$, respectively. The gray dashed line and the gray dot-dashed line denote algebraic decay scaling as ${m}^{-4}$ and ${m}^{-2}$, respectively. (d) The same data as in (c) but for $N=2$ and $M=75$.}
\label{floquet_matrix}
\end{figure}

\clearpage

\beginsupplement

\newpage

\pagenumbering{arabic} 

\begin{onecolumngrid}
\begin{center}
{\bf {\large Supplementary Materials for \\``Origin and emergent features of many-body dynamical localization''}}\\
\end{center}
\end{onecolumngrid}

\maketitle

\section{Theoretical description}
It is known that the strength of the interaction for $1$D Lieb-Liniger system can be captured by~\cite{wilson2020observation}
\begin{equation}
    \gamma=\frac{m_\mathrm{A}g_{1\mathrm{D}}}{n_{1\mathrm{D}}\hbar^2},
\end{equation}
where $g_{1\mathrm{D}}$ is the strength of the effective one-dimensional contact interaction and $n_{1\mathrm{D}}$ is the particle density. Following the dimensionless transformation in Ref.~\cite{guo2023observation}, we have 
\begin{equation}
    g_{1\mathrm{D}}=\frac{m_\mathrm{A}g}{8k^3_\mathrm{L}T^2},\quad\quad
    \hbar_{\rm{eff}} = \frac{4T\hbar k_{\mathrm{L}}^2}{m_\mathrm{A}},\quad\quad
    n_{1\mathrm{D}}=\frac{2k_\mathrm{L}N}{L}.
\end{equation}
With this, the dimensionless Lieb-Liniger parameter is given by
\begin{equation}
    \gamma=\frac{gL}{N\hbar^2_{\rm{eff}}},
\end{equation}
which allows us to determine whether the system is in the strong interaction limit ($\gamma\gg1$, degenerate solved) or the weak interaction limit ($\gamma\ll1$, degenerate). In particular, in our simulations for $N=3$, the system reaches the strongly correlated regime at $g\sim0.48$ ($\gamma=1$), where the theory in the mean-field approximation becomes invalid.

\begin{figure}
\centering
\includegraphics[width=0.6\linewidth]{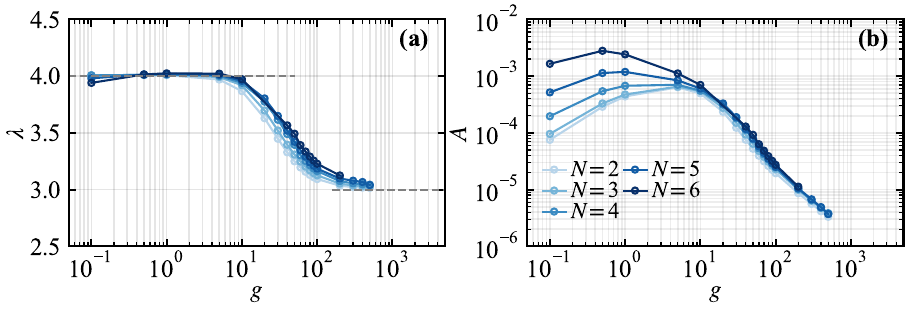}
\caption{
The decay exponent $\lambda$ (a) and the amplitude $A$ (b) as a function of $g$ for different $N$.
}
\label{Supp_unit_g}
\end{figure}

Here, we show the exponent and amplitude of the algebraic tail as a function of $g$ in Fig.~\ref{Supp_unit_g}. Different from using $\gamma$ as unit in the main text, the crossover point of $\lambda$ for different $N$ is slightly different. This also indicates that the effective interaction is affected by density.

\subsection{Derivation of the Hamiltonian in the bosonic representation}
In the language of second quantization, the general form of a two-body interacting Hamiltonian reads
\begin{equation}
    \hat{H}(t) = \int \hat{\Psi}^{\dagger}(x)\hat{h}(x,t)\hat{\Psi}(x) dx + \frac{1}{2}\iint \hat{\Psi}^{\dagger}(x)\hat{\Psi}^{\dagger}(x')U(x,x')\hat{\Psi}(x)\hat{\Psi}(x')dxdx',
\label{general_form}
\end{equation}
where $\hat{h}(x)$ and $U(x,x')$ represent the one-body Hamiltonian and the two-body interacting potential, respectively, and $\hat{\Psi}(x)$ is the field operator. For the many-body quantum kicked rotor model, we have
\begin{equation}
\begin{aligned}
    \hat{h}(x,t) = -\frac{\hbar^2_{\mathrm{eff}}}{2}\frac{\partial^2}{\partial x^2}+K\mathrm{cos}(x)\sum_{n}\delta(t-n), \quad\quad
    U(x,x') = g\delta(x-x').
\end{aligned}
\label{specific_form}
\end{equation}
By expanding the field operator $\hat{\Psi}(x)$ in the complete single-particle momentum basis $\psi_m(x)=\big({1}/{\sqrt{L}}\big) e^{-imx}$, we get
\begin{equation}
\begin{aligned}
    \hat{\Psi}(x)=\sum^{\infty}_{m=-\infty} \psi_m(x)\hat{b}_m,\quad\quad
    \hat{\Psi}^{\dagger}(x)=\sum^{\infty}_{m=-\infty} \psi^{*}_m(x)\hat{b}^{\dagger}_m.
\end{aligned}
\label{field_expansion}
\end{equation}
Intuitively, the total Hilbert space is expanded by the tensor products of all possible single-particle states $\bigotimes_{i=1}^{N}\lvert m_i\rangle$. Considering the nature of bosons, the system is fully symmetric under permutations. The basis of Hilbert space is then restricted to
\begin{equation}
    \lvert n_{-\infty}\cdots n_{m}\cdots n_{+\infty}\rangle=\sqrt{\frac{\prod_{m}n_m!}{N!}}\sum_{p}\hat{P}\lvert m_1m_2\cdots m_N\rangle,
\end{equation}
where $\hat{n}_m=\hat{b}^\dagger_m\hat{b}_m$ is the particle number occupying on state $m$ and $\hat{P}$ denotes the permutation operator.
In this representation, we have the following rules:
\begin{equation}
\begin{aligned}
        \hat{b}^\dagger_m\lvert n_{-\infty}\cdots n_{m}\cdots n_{+\infty}\rangle&=\sqrt{n_m+1}\ \lvert n_{-\infty}\cdots n_{m}+1\cdots n_{+\infty}\rangle, \\
        \hat{b}_m\lvert n_{-\infty}\cdots n_{m}\cdots n_{+\infty}\rangle&=\sqrt{n_m}\ \lvert n_{-\infty}\cdots n_{m}-1\cdots n_{+\infty}\rangle.
\end{aligned}
\end{equation}
Substituting Eqs.~(\ref{specific_form})~and~(\ref{field_expansion}) into Eq.~(\ref{general_form}), we have
\begin{equation}
\begin{aligned}
    \hat{H}(t) &= \int \sum^{\infty}_{m,m'}\frac{\hbar^2_{\mathrm{eff}}m^2}{2}\psi^{*}_m(x)\psi_{m'}(x)\hat{b}^{\dagger}_m\hat{b}_{m'}dx + \int \sum^{\infty}_{m,m'} \frac{K}{2}\left[\psi^{*}_{m+1}(x)\psi_{m'}(x)+\psi^{*}_m(x)\psi_{m'+1}(x)\right]\hat{b}^{\dagger}_m\hat{b}_{m'}dx \sum_{n}\delta(t-n) \\
    &+ \int \sum^{\infty}_{m,n,p,q}\frac{g}{2L}\psi^{*}_{m+n}(x)\psi_{p+q}(x)\hat{b}^{\dagger}_m\hat{b}^{\dagger}_n\hat{b}_p\hat{b}_q dx.
\end{aligned}
\end{equation}
Utilizing the orthogonality and completeness of the basis that $\int \psi^{*}_m(x)\psi_{m'}(x)dx=\delta_{m,m'}$, the total Hamiltonian can be expressed as
\begin{equation}
    \hat{H}(t) = \sum_{m=-\infty}^{\infty} \frac{\hbar_\mathrm{eff}^2m^2}{2}\hat{b}^\dagger_m\hat{b}_m + \frac{g}{2L}\sum^{\infty}_{m,n,p,q}\hat{b}^\dagger_{m}\hat{b}^\dagger_{n}\hat{b}_{p}\hat{b}_{q}\delta_{m+n,p+q}
    +\frac{K}{2}\left(\sum_{m=-\infty}^{\infty}\hat{b}^\dagger_m\hat{b}_{m+1} + \mathrm{h.c.}\right)\sum_{n}\delta(t-n).
\end{equation}

\subsection{Details of the lattice mapping and qualitative analysis}

\subsubsection{General mapping to the lattice model} 

According to Floquet's theorem, the Floquet eigenstates read
\begin{equation}
    \hat{U}|\phi_{\alpha}\rangle = e^{-i\frac{\theta_{\alpha} T}{\hbar_{\rm{eff}}}}|\phi_{\alpha}\rangle,
\label{floquet_states}
\end{equation}
where $0<\theta_{\alpha}\leq2\pi\hbar_{\rm{eff}}$ is the corresponding quasienergy. Firstly, we introduce the following expression
\begin{equation}\label{eq:K_expansion}
    e^{-i\frac{\hat{H}_\mathrm{K}}{\hbar_{\rm{eff}}}} = \sum_{\mathcal{K}} \frac{1-i\tan\left(\frac{E_{\mathcal{K}}}{2\hbar_{\rm{eff}}}\right)}{1+i\tan\left(\frac{E_{\mathcal{K}}}{2\hbar_{\rm{eff}}}\right)} \lvert \mathcal{K}\rangle\langle \mathcal{K}\rvert
\end{equation}
with spectral decomposition $\hat{H}_\mathrm{K}=\sum_{\mathcal{K}} E_{\mathcal{K}}\lvert \mathcal{K}\rangle\langle\mathcal{K}\rvert$. 
Similarly, the static part can be written as
\begin{equation}\label{eq:I_expansion}
    e^{-i\frac{T}{\hbar_{\rm{eff}}}\left(\hat{H}_\mathrm{I}-\theta_{\alpha}\right)} = \sum_{\mathcal{I}} \frac{1-i\tan\left(\frac{E_{\mathcal{I}}-\theta_{\alpha}}{2\hbar_{\rm{eff}}}T\right)}{1+i\tan\left(\frac{(E_{\mathcal{I}}-\theta_{\alpha})}{2\hbar_{\rm{eff}}}T\right)} \lvert \mathcal{I}\rangle\langle \mathcal{I}\rvert,
\end{equation}
where $\lvert \mathcal{I}\rangle$ and $E_{\mathcal{I}}$ denote the eigenstate and eigenenergy of $\hat{H}_\mathrm{I}$, respectively. Then, one can make the following expansion in the interacting eigenbasis $\{\lvert \mathcal{I}\rangle\}$~\cite{delande2009metalinsulator}
\begin{equation}\label{eq:interaction_expansion}
    \sum_{\mathcal{K}} \frac{1}{1+i\tan\left(\frac{E_{\mathcal{K}}}{2\hbar_{\rm{eff}}}\right)} \lvert \mathcal{K}\rangle\langle \mathcal{K}\rvert \phi_{\alpha}\rangle=\sum_{\mathcal{I}} \Pi_{\mathcal{I}}\lvert \mathcal{I}\rangle.
\end{equation}
Substituting Eqs.~(\ref{eq:K_expansion}), (\ref{eq:I_expansion}), and (\ref{eq:interaction_expansion}) into Eq.~(\ref{floquet_states}), we have the discrete form of the Schr\"{o}dinger equation as
\begin{equation}
    V_{\mathcal{I}}\Pi_{\mathcal{I}} + \sum_{\mathcal{I}\neq \mathcal{I'}}W_{\mathcal{II'}}\Pi_{\mathcal{I'}} = -W_{{\mathcal{II}}}\Pi_{\mathcal{I}}, 
\label{discrete_eq}
\end{equation}
where $V_{\mathcal{I}}=\tan[(E_{\mathcal{I}}-\theta_{\alpha})T/2\hbar_{\rm{eff}}]$ and $W_{{\mathcal{II'}}}=\sum_{\mathcal{K}} \tan(E_{\mathcal{K}}/2\hbar_{\rm{eff}})\langle \mathcal{I}\lvert\mathcal{K}\rangle\langle\mathcal{K}\rvert\mathcal{I'}\rangle$ are the on-site potential and off-diagonal hopping, respectively.

\subsubsection{The perturbative case} 

For $N>1$ with $g\neq 0$, we divide $\hat{H}_\mathrm{I}$ into three parts in the Fock basis: the free part $\hat{H}_0$, the $g-$dependent diagonal part $\hat{H}_\mathrm{ID}$ and off-diagonal part $\hat{H}_\mathrm{IOD}$. They are respectively given by
\begin{equation}
\begin{aligned}
    \hat{H}_\mathrm{I}=\hat{H}_0 + g\hat{H}_\mathrm{ID}+g\hat{H}_\mathrm{IOD}, 
\end{aligned}
\end{equation}
where 
\begin{equation}
    \hat{H}_0= \frac{1}{2}\sum_{m=-\infty}^{\infty} \hbar_\mathrm{eff}^2m^2\hat{n}_m, \quad\quad
    \hat{H}_\mathrm{ID}= \frac{1}{2L}\sum_{p,q}\hat{n}_{p}\hat{n}_{q}, \quad\quad
    \hat{H}_\mathrm{IOD}=\frac{1}{2L}\sum_{m\neq p, m\neq q}\hat{b}^\dagger_{m}\hat{b}^\dagger_{n}\hat{b}_{p}\hat{b}_{q}\delta_{m+n,p+q}.
\end{equation}
There are two scenarios: \\
(i) If $\hat{H}_\mathrm{IOD}$ is ignored, then the on-site term $V_{\mathcal{I}}$ becomes $g$-dependent nonlinear disorder while $W_{{\mathcal{II'}}}$ still decays rapidly. In such a nonlinear lattice model, localization is expected. \\
(ii) Considering small $g$ and non-degenerate case, one can apply the perturbation theory to expand the eigenenergy as $E_{\mathcal{I}}=E_{\mathcal{I}}^{(0)}+gE_{\mathcal{I}}^{(1)}+g^2E_{\mathcal{I}}^{(2)}$ and eigenfunction as $\rvert\mathcal{I}\rangle=\rvert\mathcal{I}^{(0)}\rangle+g\rvert\mathcal{I}^{(1)}\rangle$. Here, $\rvert\mathcal{I}^{(0)}\rangle$ is the eigenstate of $\hat{H}_0$. In this scenario, we have
\begin{equation}
\begin{aligned}
        E_{\mathcal{I}}^{(1)}=\langle\mathcal{I}^{(0)}\lvert \hat{H}_\mathrm{ID}\rvert\mathcal{I}^{(0)}\rangle, \quad\quad
        E_{\mathcal{I}}^{(2)}= \sum_{\mathcal{I'}\neq \mathcal{I}}\frac{\lvert\langle\mathcal{I'}^{(0)}\lvert \hat{H}_\mathrm{IOD}\rvert\mathcal{I}^{(0)}\rangle\rvert^2}{E_{\mathcal{I}}^{(0)}-E_{\mathcal{I’}}^{(0)}}, \quad\quad
        \rvert\mathcal{I}^{(1)}\rangle=\sum_{\mathcal{I'}\neq \mathcal{I}}\frac{\langle\mathcal{I'}^{(0)}\lvert \hat{H}_\mathrm{IOD}\rvert\mathcal{I}^{(0)}\rangle}{E_{\mathcal{I}}^{(0)}-E_{\mathcal{I’}}^{(0)}}\rvert\mathcal{I'}^{(0)}\rangle.
\end{aligned}
\end{equation}
Substituing the equations above into Eq.~(\ref{discrete_eq}), we have the diagonal term as
\begin{equation}
V_{\mathcal{I}}=\tan\left(\frac{ E_{\mathcal{I}}^{(0)}+gE_{\mathcal{I}}^{(1)}+g^2E_{\mathcal{I}}^{(2)}-\theta_\alpha}{2\hbar_\mathrm{eff}}T\right),
\end{equation}
and the off-diagonal term as
\begin{equation}
    W_{{\mathcal{II'}}}=W_{{\mathcal{II'}}}^{(0)}
    +g\sum_{\mathcal{I''}\neq \mathcal{I'}}\frac{\langle\mathcal{I''}^{(0)}\lvert \hat{H}_\mathrm{IOD}\rvert\mathcal{I'}^{(0)}\rangle}{E_{\mathcal{I'}}^{(0)}-E_{\mathcal{I''}}^{(0)}}W_{{\mathcal{II''}}}^{(0)}
    +g\sum_{\mathcal{I''}\neq \mathcal{I}}\frac{\langle\mathcal{I}^{(0)}\lvert \hat{H}_\mathrm{IOD}\rvert\mathcal{I''}^{(0)}\rangle}{E_{\mathcal{I}}^{(0)}-E_{\mathcal{I''}}^{(0)}}W_{{\mathcal{I''I'}}}^{(0)}
    +O(g^2),
\label{perturbative_hopping}
\end{equation}
where the unperturbed hopping is $W_{{\mathcal{II'}}}^{(0)}=\sum_{\mathcal{K}} \tan(E_{\mathcal{K}}/2\hbar_{\rm{eff}})\langle \mathcal{I}^{(0)}\lvert\mathcal{K}\rangle\langle\mathcal{K}\rvert\mathcal{I'}^{(0)}\rangle$. Thus, the off-diagonal hopping consists of the hopping of free bosons (FB) and $g$-dependent corrections.

\subsubsection{The Tonks-Girardeau limit} 

Next, we consider the non-perturbative case with $g\rightarrow\infty$. In this case, the strong local repulsion leads to the situation that different bosons cannot simultaneously occupy the same position $x$, i.e., the hard-core bosons or Tonks-Girardeau (TG) gas. Utilizing the technique of the Bose-Fermi mapping, the eigenstate of the TG gas straightforwardly takes the form of a Slater determinant and the TG gas have similar spectrum as free fermions. Thus, we have~\cite{konik2020MBDL,rigol2011coldatomreview}
\begin{equation}
    \lvert\mathcal{I}\rangle =\int dx^{N} \mathcal{A}\, \mathrm{det}\left[\psi^{\mathcal{I}}_{m_k}(x_k)\right]\prod \limits_{i=1}^N \hat{b}^\dagger_{x_i}\lvert 0\rangle,\quad\quad E_{\mathcal{I}}=\frac{1}{2}\sum_{k=1}^N\hbar^2_{\mathrm{eff}}m_k^2,
\label{TG_limit}
\end{equation}
where $\mathcal{A}=\prod_{1\leq i<j\leq N}\mathrm{sgn}(x_i-x_j)$ is the antisymmetrizer
which guarantees the permutation symmetry of the bosonic wave function and $\psi^{\mathcal{I}}_{m_k}(x_k)$ are a set of free-fermion eigenstates with quasi-momentum $m_k$. Thus, the fermionized quasi particles form a shifted Fermi sea. Here, $\hat{b}^\dagger_{x_i}$ is the creation operator of the $i-$th boson obeying to the relation $\big[\hat{b}_{x_i},\hat{b}^\dagger_{x'_i}\big]=0$ for $x_i\neq x'_i$. The on-site constraints are $\hat{b}^\dagger_{x_i}\hat{b}^\dagger_{x_i}=\hat{b}_{x_i}\hat{b}_{x_i}=0$ and $\big\{\hat{b}_{x_i},\hat{b}^\dagger_{x_i}\big\}=1$. Substituting Eq.~(\ref{TG_limit}) into Eq.~(\ref{discrete_eq}), we have
\begin{equation}
    W_{{\mathcal{II'}}}=\int dx^{N}\tan\left[\sum_i^N K\cos(x_i)/2\hbar_{\mathrm{eff}}\right]\mathcal{A}\,\mathrm{det}\left[\psi^{\mathcal{I}}_{m_k}(x_k)\right]^{*}\mathcal{A}\,\mathrm{det}\left[\psi^{\mathcal{I'}}_{m_k}(x_k)\right], \quad\quad
    V_{\mathcal{I}}=\tan\left[\left(\sum_{k=1}^N\frac{\hbar^2_{\mathrm{eff}}m^2_k}{2}-\theta_{\alpha}\right)\frac{T}{2\hbar_\mathrm{eff}}\right].
\label{TG_hopping}
\end{equation}
When $\mathcal{A}^2=1$, the off-diagonal hopping reduces to the hopping of free fermions (FF), denoted by $W^{\mathrm{FF}}_{{\mathcal{II'}}}$.

\subsubsection{Dynamical localization in the CM momentum}
It is noteworthy that the LL Hamiltonian $\hat{H}_{\mathrm{I}}$ commutes with the total momentum operator $\hat{P}$
\begin{equation}
\left[\hat{H}_{\mathrm{I}}, \hat{P}\right]=0,\quad  \hat{P}=\sum_i^N \hat{p}_i.
\end{equation}
Thus the LL eigenstates can be factorized as two parts:
\begin{equation}
\lvert\mathcal{I}\rangle =\lvert \psi_Q\rangle\otimes\lvert \psi_{q_1,q_2,\cdots}\rangle
\end{equation}
where $\lvert \psi_Q\rangle$ is a plane wave with the total momentum $Q$, the rest wave function $\lvert \psi_{q_1,q_2,\cdots}\rangle$ is about the relative momenta $q_1,q_2,\cdots$. The off-diagonal term $W_{{\mathcal{II'}}}$ can be simplified as follows:
\begin{equation}
\begin{aligned}
W_{{\mathcal{II'}}}&=\langle \mathcal{I}\lvert\tan{\left(\hat{H}_{\mathrm{K}}/2\hbar_{\mathrm{eff}}\right)}\lvert \mathcal{I'}\rangle \\
&= \langle\psi_{q_1,q_2,\cdots}\lvert\langle \psi_Q\lvert\tan{\left(\hat{H}_{\mathrm{K}}/2\hbar_{\mathrm{eff}}\right)}\lvert \psi_{Q'}\rangle\lvert \psi_{q'_1,q'_2,\cdots}\rangle \\
&=\int dx_{\mathrm{r}_1}dx_{\mathrm{r}_2}\cdots dx_{\mathrm{r}_{N}} \psi_{q_1,q_2,\cdots}{\psi}^*_{q'_1,q'_2,\cdots}\int dx_{\mathrm{c}} \tan{\left[\sum_i^N K\cos{(x_i)}/2\hbar_{\mathrm{eff}}\right]}e^{-i(Q-Q')x_{\mathrm{c}}}
\end{aligned}
\end{equation}
where $x_{\mathrm{c}}, x_{\mathrm{r}}, x_i$ denote the CM, relative and total coordinate. Noticing the identity that
\begin{equation}
\begin{aligned}
\sum_i^N \cos{(x_i)}&=\sum_i^{N}
\cos{(x_{\mathrm{c}}+x_{\mathrm{r}_i}/N)}\\
&=\cos{(x_{\mathrm{c}})}\sum_i^{N}\cos{(x_{\mathrm{r}_i}/N)}-\sin{(x_{\mathrm{c}})}\sum_i^{N}\sin{(x_{\mathrm{r}_i}/N)}\\
&=\sqrt{{\left[\sum_i^{N}\cos{(x_{\mathrm{r}_i}/N)}\right]}^2+{\left[\sum_i^{N}\sin{(x_{\mathrm{r}_i}/N)}\right]}^2}\cos{\left[x_{\mathrm{c}}+\phi(x_{\mathrm{r}_1},x_{\mathrm{r}_2},\cdots)\right]},
\end{aligned}
\end{equation}
where we introduce a phase factor $\phi(x_{\mathrm{r}_1},x_{\mathrm{r}_2},\cdots)$ about the relative coordinate 
\begin{equation}
\cos{[\phi(x_{\mathrm{r}_1},x_{\mathrm{r}_2},\cdots)]}=\frac{{\sum_i^{N}\cos{(x_{\mathrm{r}_i}/N)}}}{\sqrt{{\left[\sum_i^{N}\cos{(x_{\mathrm{r}_i}/N)}\right]}^2+{\left[\sum_i^{N}\sin{(x_{\mathrm{r}_i}/N)}\right]}^2}}.
\end{equation}
Thus, we have 
\begin{equation}
W_{{\mathcal{II'}}}=\int dx_{\mathrm{r}_1}dx_{\mathrm{r}_2}\cdots dx_{\mathrm{r}_{N}} \psi_{q_1,q_2,\cdots}{\psi}^*_{q'_1,q'_2,\cdots}W_{QQ'}(x_{\mathrm{r}_1},x_{\mathrm{r}_2},\cdots)
\end{equation}
with
\begin{equation}
W_{QQ'}(x_{\mathrm{r}_1},x_{\mathrm{r}_2},\cdots)=\int dx_{\mathrm{c}} \tan\left\{ K\sqrt{{\left[\sum_i^{N}\cos{(x_{\mathrm{r}_i}/N)}\right]}^2+{\left[\sum_i^{N}\sin{(x_{\mathrm{r}_i}/N)}\right]}^2}\cos{\left[x_{\mathrm{c}}+\phi(x_{\mathrm{r}_1},x_{\mathrm{r}_2},\cdots)\right]}/2\hbar_{\mathrm{eff}}\right\}e^{-i(Q-Q')x_{\mathrm{c}}}.
\end{equation}
If we fix all the relative coordinate $x_{\mathrm{r}_1},x_{\mathrm{r}_2},\cdots$, the phase factor $\phi(x_{\mathrm{r}_1},x_{\mathrm{r}_2},\cdots)$ is a constant. Then $W_{QQ'}$ has the same form as that of the single-particle QKR, which is known to decay exponentially fast. Therefore, generally for any number of particles, we expect DL always holds in the CM momentum.

\subsection{Two-particle case in the lattice mapping}

We consider two bosons, with positions $(x_1,x_2)$ and treat them in a CM and relative coordinates frame ($x_\mathrm{c}=(x_1+x_2)/2$, $x_{\mathrm{r}}=x_1-x_2$). Then, the eigenfunction of the Lieb-Liniger model $\hat{H}_{\mathcal{I}}$ reads~\cite{qin2017interacting}
\begin{equation}
\psi^{q}_{Q}(x_\mathrm{c},x_{\mathrm{r}}) =\left\{
\begin{aligned}
& \sqrt{\frac{1}{\pi}}B^q_Q \cos{[qx_{\mathrm{r}}-(q+Q)\pi]}e^{2iQx_\mathrm{c}}, & & \mathrm{if}\,0\leq x_{\mathrm{r}}\leq 2\pi, \\
& \sqrt{\frac{1}{\pi}}B^q_Q \cos{[qx_{\mathrm{r}}+(q+Q)\pi]}e^{2iQx_\mathrm{c}}, & & \mathrm{if} -2\pi\leq x_{\mathrm{r}}\leq 0,
\end{aligned}
\right.
\end{equation}
where $B^q_Q={\left[{8\pi+(4/q)\sin{(2q\pi)}\cos{(2Q\pi)}}\right]}^{-1/2}$ and $2(q+Q)\pi=\pi-2\arctan(2qA_g)$ with an inverse interaction strength $A_g=\hbar^2_{\mathrm{eff}}/g$. $Q$ is the CM momentum with $Q=0,\pm1/2,\pm1,\pm3/2,\cdots$, and $q$ is the relative momentum. Thus, the eigenenergy is given by $E^q_Q=\hbar^2_{\mathrm{eff}}\left(Q^2+q^2\right)$. 

Having obtained the eigenfunction, the off-diagonal term $W^{qr}_{QR}$ in the lattice mapping is given by
\begin{equation}
    W^{qq'}_{QQ'}=2\iint {\psi^{q}_{Q}(x_\mathrm{c},x_{\mathrm{r}})}\tan{\left[\frac{K}{\hbar_{\mathrm{eff}}}\cos{x_\mathrm{c}}\cos{\frac{x_{\mathrm{r}}}{2}}\right]}{\psi^{q'}_{Q'}(x_\mathrm{c},x_{\mathrm{r}})}^{*} dx_\mathrm{c} dx_{\mathrm{r}}. \label{hopping_eq}
\end{equation}
Then, utilizing the periodicity of function, we have a simpler expression as
\begin{equation}
    W^{qq'}_{QQ'}=\frac{1}{\pi}B^q_QB^{q'}_{Q'} \int_0^{2\pi}\Big\{\cos{[\phi_{+}(x_{\mathrm{r}})]}+\cos{[\phi_{-}(x_{\mathrm{r}})]}\Big\} \mathcal{B}_{Q-Q'}(x_{\mathrm{r}})dx_{\mathrm{r}},
\end{equation}
where 
\begin{equation}
\begin{aligned}
    \mathcal{B}_{Q-Q'}(x_{\mathrm{r}})&=2\int_0^{2\pi}\tan{\left[\frac{K}{\hbar_{\mathrm{eff}}}\cos{x_\mathrm{c}}\cos{\frac{x_{\mathrm{r}}}{2}}\right]}e^{2i(Q-Q')x_\mathrm{c}}dx_\mathrm{c},\\
    \phi_{\pm}(x_{\mathrm{r}})&=(q\pm q')x_{\mathrm{r}}-(q\pm q'+Q\pm Q')\pi.
\end{aligned}
\end{equation}
After calculating a series of integrals, we have
\begin{equation}
    W^{qq'}_{QQ'}=\frac{1}{\pi}B^q_QB^{q'}_{Q'}\sum_{s=\pm}\Big[\mathcal{C}_{1}^{s}+\mathcal{C}_{2}^{s}+\mathcal{C}_{3}^{s}+\cdots+\mathcal{C}_{n}^{s}+\mathcal{D}_{n}^{s}\Big],
\end{equation}
where
\begin{equation}
\begin{aligned}
    \mathcal{C}_{2i-1}^{\pm}&=\frac{(-1)^{i+1}\mathcal{B}_{Q-Q'}^{(2i-2)}(x_{\mathrm{r}})\sin{[\phi_{\pm}(x_{\mathrm{r}})]}}{(q\pm q')^{2i-1}}\Bigg|^{2\pi}_0,\quad\quad\quad\ \ \ \
    \mathcal{C}_{2i}^{\pm}=\frac{(-1)^{i+1}\mathcal{B}_{Q-Q'}^{(2i-1)}(x_{\mathrm{r}})\cos{[\phi_{\pm}(x_{\mathrm{r}})]}}{(q\pm q')^{2i}}\Bigg|^{2\pi}_0,\\
    \mathcal{D}_{2i-1}^{\pm}&=\frac{(-1)^{i}\int_0^{2\pi}\mathcal{B}_{Q-Q'}^{(2i-1)}(x_{\mathrm{r}})\sin{[\phi_{\pm}(x_{\mathrm{r}})]}dx_{\mathrm{r}}}{(q\pm q')^{2i-1}},\quad\quad
    \mathcal{D}_{2i}^{\pm}=\frac{(-1)^{i}\int_0^{2\pi}\mathcal{B}_{Q-Q'}^{(2i)}(x_{\mathrm{r}})\cos{[\phi_{\pm}(x_{\mathrm{r}})]}dx_{\mathrm{r}}}{(q\pm q')^{2i}},
\end{aligned}
\end{equation}
with $\mathcal{B}_{Q-Q'}^{(n)}=d^n\mathcal{B}_{Q-Q'}(x_{\mathrm{r}})/dx_{\mathrm{r}}^n$. Since $\mathcal{B}_{Q-Q'}(x_{\mathrm{r}})$ is an even function with a period of $4\pi$, we have $\mathcal{B}_{Q-Q'}^{(2i-1)}(0)=\mathcal{B}_{Q-Q'}^{(2i-1)}(2\pi)=0$, implying $\mathcal{C}_{2i}^{\pm}=0$. Then, considering the leading order $C_1^{\pm}+C_3^{\pm}$ and $\mathcal{B}_{Q-Q'}(2\pi)=-\mathcal{B}_{Q-Q'}(0), \mathcal{B}_{Q-Q'}^{(2)}(2\pi)=-\mathcal{B}_{Q-Q'}^{(2)}(0)$,  we have
\begin{equation}
\begin{aligned}
    \mathcal{C}_1^{+}+\mathcal{C}_1^{-}&=2\mathcal{B}_{Q-Q'}(0)\frac{\sin{[(Q+Q')\pi]}\cos{[(q+q')\pi]}}{q+q'}+2\mathcal{B}_{Q-Q'}(0)\frac{\sin{[(Q-Q')\pi]}\cos{[(q-q')\pi]}}{q-q'},\\
    \mathcal{C}_3^{+}+\mathcal{C}_3^{-}&=-2\mathcal{B}_{Q-Q'}^{(2)}(0)\frac{\sin{[(Q+Q')\pi]}\cos{[(q+q')\pi]}}{(q+q')^3}-2\mathcal{B}_{Q-Q'}^{(2)}(0)\frac{\sin{[(Q-Q')\pi]}\cos{[(q-q')\pi]}}{(q-q')^3}.
\end{aligned}
\end{equation}
We note that if $Q$ and $Q'$ are integers or half-integers, the above equations equal to $0$. 

\begin{figure}
\centering
\includegraphics[width=0.5\linewidth]{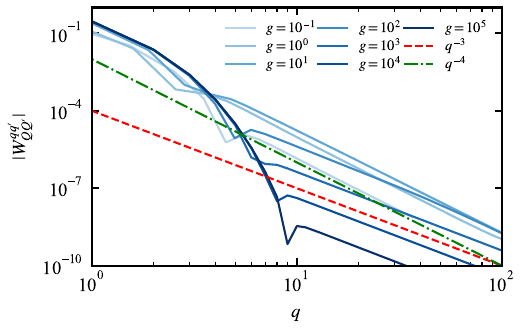}
\caption{
The off-diagonal term $|W_{QQ'}^{qq'}|$ from Eq.~(\ref{hopping_eq}) as a function of relative momentum $q$ for different interaction strengths. The red and green dashed lines denote algebraic decay as $q^{-3}$ and $q^{-4}$, respectively. The kick strength is $K=1$ and here we set $Q'=0, Q=0.5$.
}
\label{Supp_two_particle_mapping}
\end{figure}

Next, considering that $Q$ is a half-integer, $Q'$ is an integer, and $\sin{[(Q+Q')\pi]}=1$, we have
\begin{equation}
    \mathcal{C}_1^{+}+\mathcal{C}_1^{-}=2\mathcal{B}_{Q-Q'}(0)\frac{2q\cos{(q\pi)}\cos{(q'\pi)}+2q'\sin{(q\pi)}\sin{(q'\pi)}}{q^2-{q'}^2},
\end{equation}
where 
\begin{equation}
    \mathcal{B}_{Q-Q'}(0)=2\int_0^{2\pi}\tan{\left[\frac{K}{\hbar_{\mathrm{eff}}}\cos{x_\mathrm{c}}\right]}e^{2i(Q-Q')x_\mathrm{c}}dx_\mathrm{c}.
\end{equation}
Using $2(q+Q)\pi=\pi-2\arctan(2qA_g)$, $2q/\tan{(q\pi)}=-1/A_g$, and $ 2q'\tan{(q'\pi)}=1/A_g$, we have
\begin{equation}
    \mathcal{C}_1^{+}+\mathcal{C}_1^{-}=0.
\end{equation}
Considering $q\gg q'$, we also have
\begin{equation}
    \mathcal{C}_3^{+}+\mathcal{C}_3^{-}=-2\mathcal{B}_{Q-Q'}^{(2)}(0)\frac{2\cos{(q\pi)}\cos{(q'\pi)}}{q^3}.
\end{equation}
Then, since  
\begin{equation}
    \cos{(q\pi)}=\frac{\pm1}{\sqrt{[1+(2qA_g)^2]}},\quad\quad
    \cos{(r\pi)}=\frac{\pm2q'A_g}{\sqrt{[1+(2q'A_g)^2]}},
\end{equation}
we have
\begin{equation}
    \mathcal{C}_3^{+}+\mathcal{C}_3^{-}=-4\mathcal{B}_{Q-Q'}^{(2)}(0)\frac{\pm1}{\sqrt{[1+(2qA_g)^2]}}\frac{\pm2q'A_g}{\sqrt{[1+(2q'A_g)^2]}}\frac{1}{q^3}.
\end{equation}
Considering large $q$,  $B^q_Q$ and $B^{q'}_{Q'}\sim {\big[\sqrt{8\pi}\big]}^{-1}$, we arrive at
\begin{equation}
    \left\lvert W^{qq'}_{QQ'}\right\rvert\sim\bigg|\frac{\mathcal{B}_{Q-Q'}^{(2)}(0)}{2\pi^2} \frac{1}{\sqrt{[1+(2qA_g)^2]}}\frac{2q'A_g}{\sqrt{[1+(2q'A_g)^2]}}\frac{1}{q^3}\bigg|,
\end{equation}
where 
\begin{equation}
\begin{aligned}
    \mathcal{B}_{Q-Q'}^{(2)}(0)&=\int_0^{2\pi}\frac{K\cos{x_\mathrm{c}}}{2\hbar_{\mathrm{eff}}\cos^2{\left[\frac{K}{\hbar_{\mathrm{eff}}}\cos{x_\mathrm{c}}\right]}}e^{2i(Q-Q')x_\mathrm{c}}dx_\mathrm{c},
\end{aligned}
\end{equation}
decays exponentially fast with $Q-Q'$. Especially, we consider fixed $q'$ and finite $q$. For weak interaction ($A_g\rightarrow\infty$), $2qA_g\gg 1$, $2q'A_g\gg 1$ one finds that 
\begin{equation}
    \left\lvert W^{qq'}_{QQ'}\right\rvert\sim\bigg|\frac{\mathcal{B}_{Q-Q'}^{(2)}(0)}{4\pi^2A_g}\frac{1}{q^4}\bigg|,
\end{equation}
whereas for the TG limit ($A_g\rightarrow 0$), we have $2qA_g\ll 1$, $2q'A_g\ll 1$, and 
\begin{equation}
   \left\lvert W^{qq'}_{QQ'}\right\rvert\sim\bigg|\frac{\mathcal{B}_{Q-Q'}^{(2)}(0)q'A_g}{\pi^2}\frac{1}{q^3}\bigg|.
\end{equation}
For integer $Q$ and half-integer $Q'$, one can arrive at the same result. Therefore, from the viewpoint of the lattice mapping, the tails of off-diagonal hopping for weak and strong interactions exhibit distinct decay behaviors. Notably, the amplitude of the tail at weak interaction contributes a lot ($\propto A_g^{-1}$), whereas it becomes smaller and even negligible with the increase of interactions ($\propto A_g$). This explains the absence of tail in Eq.~(\ref{TG_hopping}). As shown in Fig.~\ref{Supp_two_particle_mapping}, as $g$ increases, the amplitude of tail undergoes a nonmonotonic change and the exponent changes from $q^{-4}$ to $q^{-3}$.

\subsection{The off-diagonal term at the small K limit}
In the fundamental sector ($x_1\leq x_2\leq \cdots\leq x_N$) and ($m_1<m_2<\cdots<m_N$), the eigenstates $\lvert \mathcal{I}\rangle=\lvert m_1m_2\cdots m_N\rangle$ of the LL model are known as
\begin{equation}
    \Psi_{\{m\}}({x_j})=\sum_{P\in S_N}A_P\mathrm{exp}\left(i\sum_{j=1}^Nm_{P(j)}x_j\right),
\label{bethe_solutions}
\end{equation}
where $S_N$ is a permutation group of $N$ quasi-momenta $\{m_j\}$, and the coefficients are
\begin{equation}
    A_P = \frac{(-1)^{\mathrm{sgn}(P)}\prod_{j>l}\left[m_{P(j)}-m_{P(l)}-ig\right]}{\left\{L^NN!\prod_{j>l}[(m_j-m_l)^2+g^2]\right\}^{-1/2}}.
\label{bethe_coefficients}
\end{equation}
These quasi-momenta are coupled by $N$ Bethe equations
\begin{equation}
    m_j=\frac{2\pi}{L}I_j-\frac{2}{L}\sum_{l=1}^{N}\arctan{\left(\frac{m_j-m_l}{g}\right)},
\label{bethe_equations}
\end{equation}
where $I_j$ are ingeters for odd $N$ and half-integers for even $N$. For $g=0$ and $\infty$, Eq.~\eqref{bethe_solutions} reduces to the free
bosons’ wavefunction and the Slater’s determinant with an
antisymmetrizer, respectively. In the two limits, $m_j$  are equidistant momentum. Yet for finite $g$, $m_j$  are strongly coupled and unequally spaced.

We aim to give a unified off-diagonal term for any $N$. In terms of the Bethe states, the off-diagonal term reads 
\begin{equation}
    W_{\mathcal{I}\mathcal{I}'}= N!\sum_{P'P}A_{P'}^*A_P\int_{x_1\leq x_2\leq \cdots\leq x_N}d^Nx_j \tan{\Big(\frac{K}{2\hbar_{\mathrm{eff}}}\sum_j\cos{x_j}\Big)}\mathrm{exp}\left\{i\sum_jx_j[m_{P(j)}-m'_{P'(j)}]\right\}.
\end{equation}
For small $K$, $\tan{\big(\frac{K}{2\hbar_{\mathrm{eff}}}\sum_j\cos{x_j}\big)}\approx \frac{K}{2\hbar_{\mathrm{eff}}}\sum_j\cos{x_j}$, which mimics the nearest-neighbor hopping between quasi-momentum. Choosing $x_1$ as a moving boundary, we have the approximate $W_{\mathcal{I}\mathcal{I}'}$ as
\begin{equation}
    W_{\mathcal{I}\mathcal{I}'}=N!\frac{K}{2\hbar_{\mathrm{eff}}}\sum^N_{h=1}\sum_{P'P}A_{P'}^*A_PS^{h}_{{\mathcal{I}\mathcal{I}'}PP'},
\end{equation}
with 
\begin{equation}
S^{h}_{{\mathcal{I}\mathcal{I}'}PP'}=\int_0^Ldx_1\prod_{j=2}^N\int_{x_{j-1}}^Ldx_j\cos{x_h}\mathrm{exp}{\left\{i\sum^N_{j=1}x_j[m_{P(j)}-m'_{P'(j)}]\right\}},
\end{equation}
where the factor $N!$ accounts for all the equivalent permutations of the $x_j$. The above form indicates that, generally for arbitrary $N$, the kicks contribute a cos function that simply alters the difference of CM momentum between two Bethe states, as we see in the analytic form of $N=2$.

To evaluate the above integral, we introduce a new set of variables $y_j$ which satisfies
\begin{equation}
    y_1 = x_1,\quad y_j=x_j-x_{j-1}\,(j\geq2),\quad y_{N+1}=L-x_N,\quad \sum_{j=1}^{N+1}y_j=L,\quad x_l=\sum_{j=1}^l y_j.
\end{equation}
Then we introduce the following quantities describing the difference between quasi-momenta
\begin{equation}
\begin{aligned}
    \Delta m^{PP'}_l&=m_{P(l)}-m'_{P'(l)}, \\
    c^{s,PP'}_{h,j}&=\sum_{l=j}^{N}\Delta m^{PP'}_l+s\cdot H(h-j),
\end{aligned}
\end{equation}
where $s=\pm1,\,1\leq h\leq N,\, 1\leq j\leq N+1$ and $H(\cdot)$ is the Heaviside step function defined as
\begin{equation}
H(x)=\left\{
\begin{aligned}
& 1 & & \mathrm{for}\enspace x\geq0, \\
& 0 & & \mathrm{for}\enspace x<0. 
\end{aligned}
\right.
\end{equation}
Thus we have
\begin{equation}
    \mathrm{exp}\left\{i\sum_{j=1}^N\Delta m_j^{PP'}x_j\right\}=\mathrm{exp}\left\{i\sum_{j=1}^N\Delta m_j^{PP'}\sum^j_{l=1}y_l\right\}=\mathrm{exp}\left\{i\sum_{j=1}^{N+1} c^{s,PP'}_{h,j}y_j\right\},
\end{equation}
and the original integral now turns into
\begin{equation}
S^{h}_{{\mathcal{I}\mathcal{I}'}PP'}=\frac{1}{2}\sum_{s=\pm}\int_0^L \prod_{j=1}^{N+1} dy_j\delta\left(\sum_{j=1}^{N+1}y_j-L\right) \mathrm{exp}{\left\{i\sum^{N+1}_{j=1}c^{s,PP'}_{h,j} y_j\right\}}.
\end{equation}
After substituting $\delta\left(\sum_{j=1}^{N+1}y_j-L\right)=\frac{1}{2\pi}\int_{-\infty}^{\infty}\mathrm{exp}\left\{i\omega(\sum_{j=1}^{N+1}y_j-L)\right\}d\omega$ into the above integral, $S^{h}_{{\mathcal{I}\mathcal{I}'}PP'}$ reads
\begin{equation}
\begin{aligned}
    S^{h}_{{\mathcal{I}\mathcal{I}'}PP'}&=\frac{1}{4\pi}\sum_{s=\pm1}\int_{-\infty}^{\infty}\mathrm{exp}\{-i\omega L\}d\omega\prod_{j=1}^{N+1}\int_0^{L}\mathrm{exp}\left\{i(\omega+c^{s,PP'}_{h,j})y_j\right\}dy_j \\
    &= \frac{1}{4\pi}\sum_{s=\pm1}\int_{-\infty}^{\infty}\mathrm{exp}\{-i\omega L\}\frac{i^{N+1}}{\prod_{j=1}^{N+1}(\omega+c^{s,PP'}_{h,j}+i\epsilon)}d\omega,
\end{aligned}
\end{equation}
where we expand the upper bound of $y_j$ to $\infty$ and use the analytic continuation
\begin{equation}
    \int_0^{\infty}\mathrm{exp}\{i(\omega+c^{s,PP'}_{h,j})y_j\}dy_j=\lim_{\epsilon\rightarrow0^{+}}\frac{i}{\omega+c^{s,PP'}_{h,j}+i\epsilon}.
\end{equation}
By applying the residue theorem, the integral is then the sum of a series of residues
\begin{equation}
S^{h}_{{\mathcal{I}\mathcal{I}'}PP'}=\frac{i^N}{2}\sum_{s=\pm1}\sum_{l=1}^{N+1}\mathrm{Res}\left[\frac{\mathrm{exp}\{-i\omega L\}}{\prod_{j=1}^{N+1}(\omega+c^{s,PP'}_{h,j}+i\epsilon)}\right]\bigg\lvert_{w=-c^{s,PP'}_{h,l}}.
\label{S_residue}
\end{equation}
So far, we have obtained a unified off-diagonal term $W_{{\mathcal{I}\mathcal{I}'}}$ for small $K$. After solving all the Bethe equations, the corresponding coefficients and the residues, it can be used to compute the algebraic decay for any $N$. 
 
\begin{figure}
\centering
\includegraphics[width=0.8\linewidth]{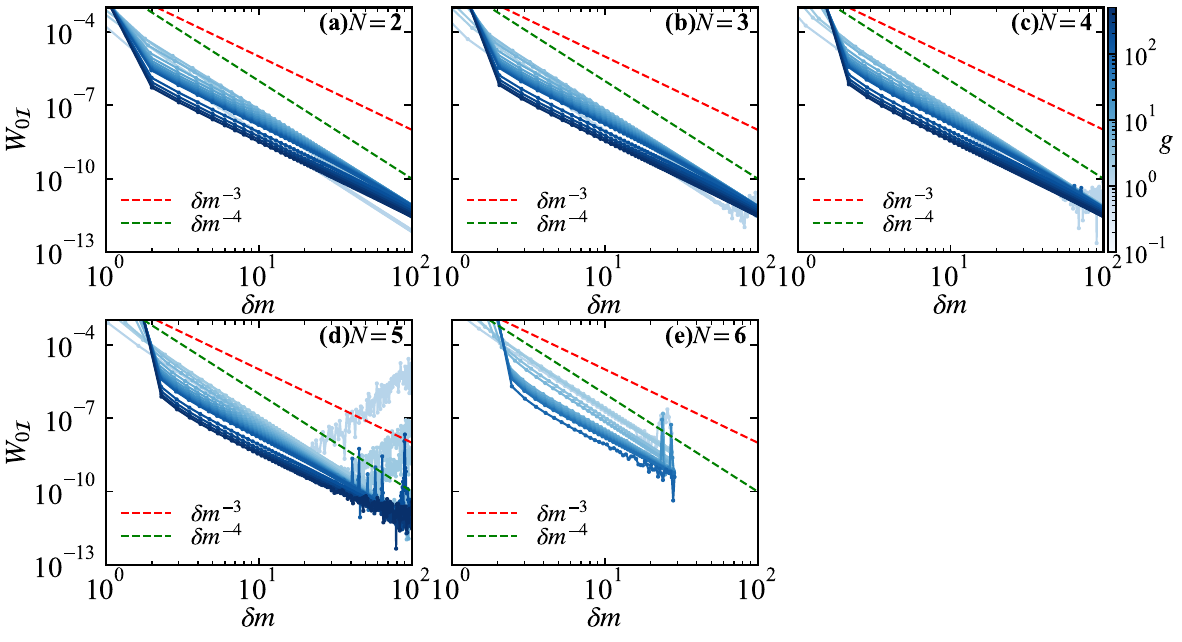}
\caption{
The approximate off-diagonal term $W_{0\mathcal{I}}$ as a function of $\delta m$ for different particle numbers $N$ and interaction strengths $g$. The colorbar denotes different interactions $g$. The red and green dashed lines denote algebraic decay $\delta m^{-3}$ and $\delta m^{-4}$, respectively.
}
\label{Supp_approx_W}
\end{figure}

In this small $K$ limit, the decay in the total momentum is approximately a $\delta-$function at $2\pi/L$, thus the off-diagonal term is non-zero only when the total momentum of two Bethe states differs at most by $2\pi/L$. This holds when $\sum_{l=1}^N \Delta m_l^{PP'}=2\pi/L$. Hence, Eq.~\eqref{S_residue} always has a second-order pole at $c_{h,1}^{s,PP'}=c_{h,N+1}^{s,PP'}$ induced by the kicks. In this case, the two Bethe states belong to different momentum sector such that $\Delta m_l^{PP'}\neq0$ for any $l$, and thus more particles just bring more first-order poles. In this sense, the algebraic features should be universal. Beyond the small $K$ limit, we just need to consider more momentum sectors. Starting from the ground state, We show the computed $W_{0\mathcal{I}}$ as a function of $\delta m$ for different $N$ in Fig.~\ref{Supp_approx_W}. $\delta m$ represents the standard deviation of the quasi-momentum of the excited state $\lvert\mathcal{I}\rangle$. We only consider two-particle excitations that $m_1$ and $m_N$ are varying with $m_1+m_N=2\pi/L$ and $\sum^{N-1}_{j=2}m_j=0$. Remarkably for all $N$ considered here, the algebraic tail always exists and it undergoes a crossover from $\delta m^{-4}$ to $\delta m^{-3}$ as the interaction $g$ increases. This indicates the universality of the algebraic features. Note that, for large $N$, the off-diagonal term elements exhibit instability at large $\delta m$. This is attributed to the large numerical errors when solving the high excited states of $N$ Bethe equations.

\begin{figure}
\centering
\includegraphics[width=0.8\linewidth]{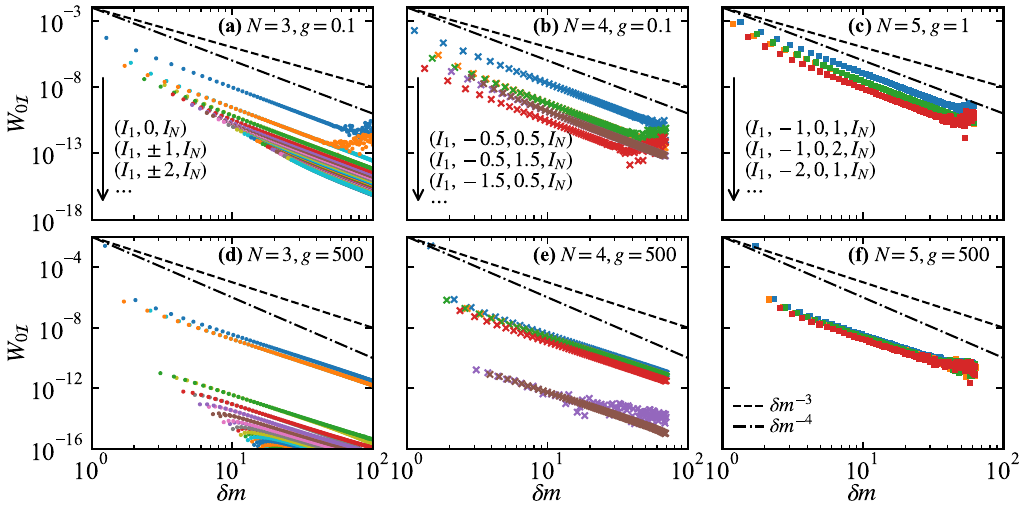}
\caption{
The approximate off-diagonal term $W_{0\mathcal{I}}$ as a function of $\delta m$ for different particle numbers $N$ and interaction strengths $g$. Different layers are obtained by varying $(I_1,I_N)$ with different sets of $(I_2,\cdots,I_{N-1})$ fixed. The black dashed and dot-dashed lines denote algebraic decay $\delta m^{-3}$ and $\delta m^{-4}$, respectively.
}
\label{Supp_diff_layers}
\end{figure}

Above, we consider the simplest case that $(m_2,\cdots,m_{N-1})$ of the excited state is the same as that of the ground state. When $(m_2,\cdots,m_{N-1})$ is varied, we notice that the algebraic tails still exist in the decay along $(m_1,m_N)$. As shown in Fig.~\ref{Supp_diff_layers}, we use quantum numbers $I_j$ to equivalently denote the corresponding quasi-momenta $m_j$. In a specific CM sector (fixed $\sum_{j=1}^N I_j=2\pi/L$), if we choose a set of fixed $(I_2,\cdots,I_{N-1})$ but vary $(I_1,I_N)$, the off-diagonal element $W_{0\mathcal{I}}$ will decay algebraically. Different sets of $(I_2,\cdots,I_{N-1})$ lead to different layers of algebraic tails. These algebraic tails from different layers simultaneously exhibit a crossover as the interaction increases. Therefore, considering the permutation symmetry of different quasi-momenta $m_j$, we conclude that the algebraic tail is typical characteristic of two-particle excitations (varying any two $m_j$'s with the others fixed). 

We can understand this universal behavior of the off-diagonal term on a general basis. At small $K$ limit and using the field operator $\Psi(x)$, the off-diagonal term reads equally
\begin{eqnarray}
    W_{\mathcal{I}\mathcal{I}'}&=&\langle\mathcal{I}\lvert \tan{\Big[\int\frac{K}{2\hbar_{\mathrm{eff}}}\cos{x}\Psi^{\dagger}(x)\Psi(x)dx\Big]}\rvert\mathcal{I}'\rangle\approx \int\frac{K}{2\hbar_{\mathrm{eff}}}\cos{x}\langle\mathcal{I}\lvert\Psi^{\dagger}(x)\Psi(x)\rvert\mathcal{I}'\rangle dx= \nonumber\\    
    &=&\frac{K}{2\hbar_{\mathrm{eff}}}\int\cos{x}\langle\mathcal{I}\lvert e^{i \hat{P}x}\Psi^{\dagger}(0)\Psi(0)e^{-i \hat{P}x}\rvert\mathcal{I}'\rangle dx=
    \frac{K}{2\hbar_{\mathrm{eff}}}\langle\mathcal{I}\lvert \Psi^{\dagger}(0)\Psi(0)\rvert\mathcal{I}'\rangle\int\cos{x}e^{i(Q-Q')x}dx,
\end{eqnarray}
where $\hat{P}$ is the total momentum operator, generator of translations, and $Q(Q')$ denote the total momentum of state $\mathcal{I}(\mathcal{I}')$. One can notice that the integral is expressing momentum conservation in the process, making sure that only states with $|Q-Q'|=1$ give a non zero matrix element. The term $\langle\mathcal{I}\lvert \Psi^{\dagger}(0)\Psi(0)\rvert\mathcal{I}'\rangle$ is nothing but the form factor of the density operator $\Psi^{\dagger}(0)\Psi(0)$ in the integrable LL model. A well-known exact form of the form factor is available~\cite{korepin1982,Slavnov1989,Caux_2007,Piroli_2015}. In the thermodynamic limit ($N, L\rightarrow\infty$) the form factor decays super-exponentially in the system size as $\sim e^{-L^2}$ if the two sates $I,I'$ correspond to thermodynamically distinct states, i.e. a macroscopic number of quasimomenta differ between the two~\cite{Essler2024prx}. A direct consequence is that the high-dimensional quantum network disconnects in most directions. Therefore, we expect that the MBDL still survive in the thermodynamic limit, at small values of $K$ and any value of $g$. 

\section{Additional numerical results}

\subsection{Additional results of the off-diagonal term in the lattice mapping}
\begin{figure}
\centering
\includegraphics[width=1\linewidth]{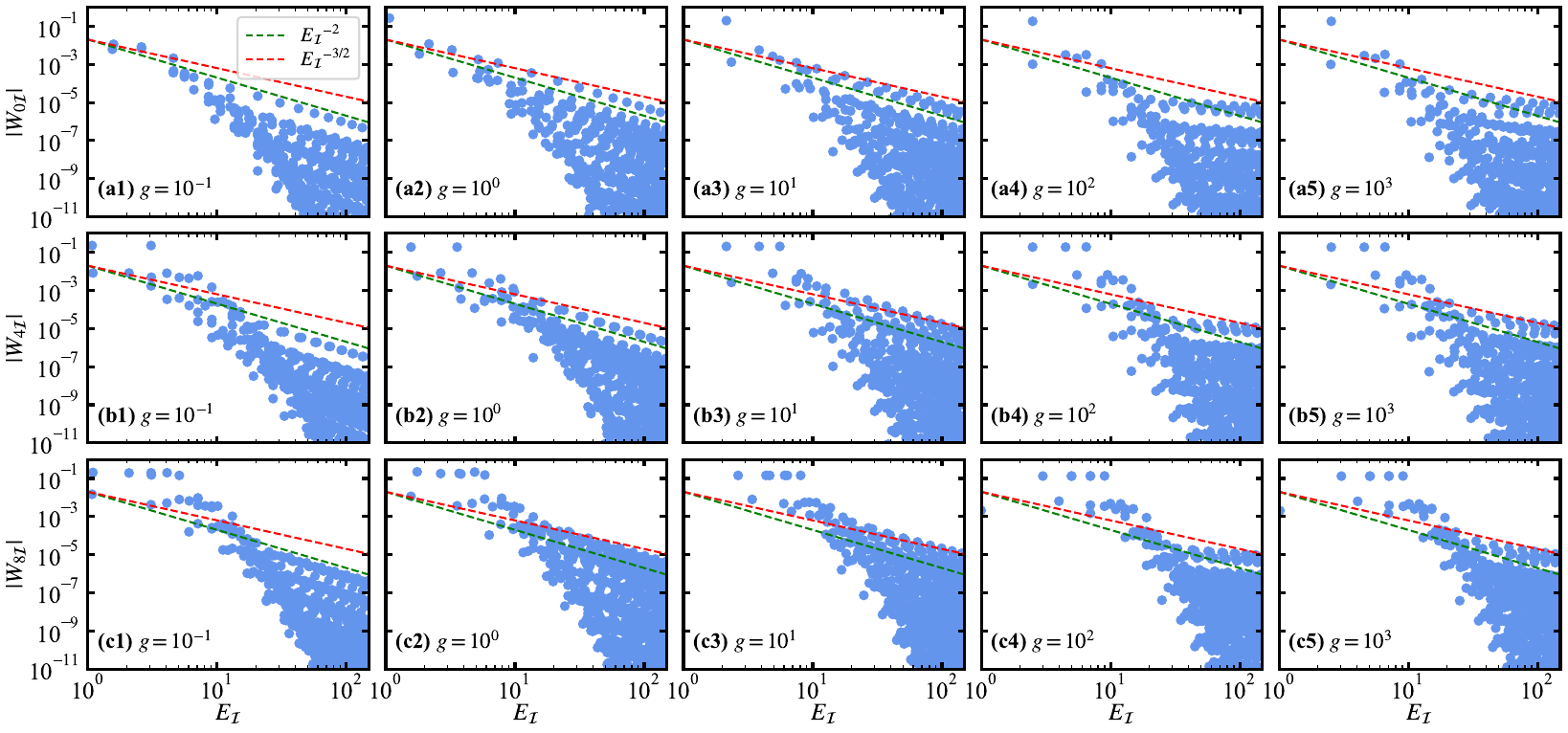}
\caption{
(a1-a5) The off-diagonal term $|W_{0\mathcal{I}}|$ as a function of energy $E_{\mathcal{I}}$ for different interactions. The red and green dashed lines denote algebraic decay as ${E_{\mathcal{I}}}^{-3/2}$ and ${E_{\mathcal{I}}}^{-2}$, respectively. The cut-off, the kick strength and the particle number are $M=35$, $K=0.5$ and $N=3$, respectively. (b1-b5) The same data as in (a1-a5) but for $|W_{4\mathcal{I}}|$. (c1-c5) The same data as in (a1-a5) but for $|W_{8\mathcal{I}}|$.
}
\label{Supp_diff_mapping}
\end{figure}
Here we show more details about the tail's crossover of the off-diagonal term $|W_{\mathcal{I}\mathcal{I'}}|$ with increasing interactions. As shown in Fig.~\ref{Supp_diff_mapping} (a1-a5), when $g\leq10$, i.e. in regime I, the off-diagonal term always exhibits a tail as ${E_{\mathcal{I}}}^{-2}$, whereas the amplitude of the tail increases with increasing $g$. As $g$ increases into regime II, the tail dramatically changes, approaching ${E_{\mathcal{I}}}^{-3/2}$ and its amplitude becomes smaller. This nonmonotonic change aligns with the two-particle prediction. If we start from an initial excited state, the crossover behavior of the tail is similar, as shown as curves of $|W_{4\mathcal{I}}|$ (b1-b5) and $|W_{8\mathcal{I}}|$ (c1-c5) in Fig.~\ref{Supp_diff_mapping}. But we can see that the decay becomes anisotropic between excited states.

\subsection{Occupation spectrum of the one-particle density matrix}

We can also characterize the deep MBDL regime (small $K$) using the one-particle density matrix (OPDM). In the Fock basis, given a Floquet eigenstate $\lvert \psi\rangle$, the OPDM is defined as~\cite{liu2021MBcritical,bera2015MBLOPDM}
\begin{equation}
    \rho_{mn} = \langle \psi\lvert \hat{b}_m^{\dagger}\hat{b}_n\rvert \psi\rangle.
\end{equation}
The natural orbitals (NOs) $\lvert\phi_{\alpha}\rangle$ and occupations $n_{\alpha}$ are obtained by diagonalizing $\rho$
\begin{equation}
    \rho\lvert\phi_{\alpha}\rangle=n_{\alpha}\lvert\phi_{\alpha}\rangle. 
\end{equation}
For interaction-free scenario, the NOs are the single-particle Anderson orbitals. In the presence of interactions, we can consider the NOs as the quasi-particle orbitals. The total occupations are $\sum_{\alpha}n_{\alpha}=N$, and $n_{\alpha}$ is similar to the quasi-local integrals of motion in the effective Hamiltonian for the MBL phase~\cite{abanin2019reveiwMBL,bera2015MBLOPDM}.
\begin{figure}
\centering
\includegraphics[width=0.65\linewidth]{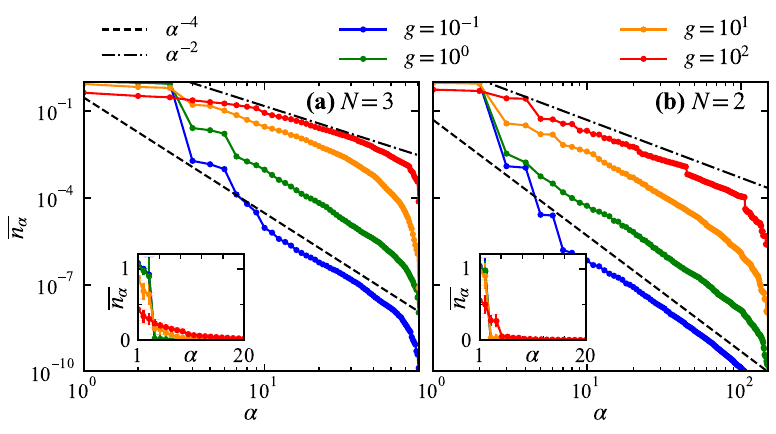}
\caption{
(a) The occupation spectrum $\overline{n}_{\alpha}$ for the particle number $N=3$ and different interaction strength $g$. The inset shows the same data but in linear scale. The cut-off and the kick strength are $M=30$ and $K=0.1$. (b) The same data as in (a) but for $N=2$ and $M=75$. The results are obtained by average over $50$ Floquet eigenstates.
}
\label{supp_occupation}
\end{figure}

We show the averaged $\overline{n}_{\alpha}$ for different $g$ ranging from $g=0.1$ to $g=100$ in Fig.~\ref{supp_occupation}. For $N=3$ and $g\leq 10$, we find that $\overline{n}_{\alpha}$ mainly distribute at the first $N$ orbitals where $\overline{n}_{\alpha}\approx 1$ (see in the inset of Fig.~\ref{supp_occupation}(a)). This is a strong localization similar to the single-particle scenario. When the system leaves regime (I) ($g=100$), the largest $\overline{n}_{\alpha}$ almost halves with a wider distribution, which still present localization. This corresponds to the extend property of the Floquet eigenstates. Remarkably different from the single-particle scenario, $\overline{n}_{\alpha}$ exhibit an algebraic tail, asymptotically falling as $\alpha^{-4}$ for $g\leq 10$ while $\alpha^{-2}$ at $g=100$. The crossover between two algebraic tails is much pronounced in the scenario of $N=2$ due to larger cut-off $M$ (Fig.~\ref{supp_occupation}(b)). These tails are induced by long-range hoppings $\hat{H}_\mathrm{IOD}=\frac{g}{2L}\sum_{m\neq p, m\neq q}\hat{b}^\dagger_{m}\hat{b}^\dagger_{n}\hat{b}_{p}\hat{b}_{q}\delta_{m+n,p+q}$ and show the coherence of the interacting bosons in Fock basis. Such a crossover also corresponds to the jump behavior of $D_{\beta}$ and $b_{\beta}$ as the interaction strength $g$ increases.

\subsection{Additional results of the hopping amplitude in Fock basis}
\begin{figure}
\centering
\includegraphics[width=1\linewidth]{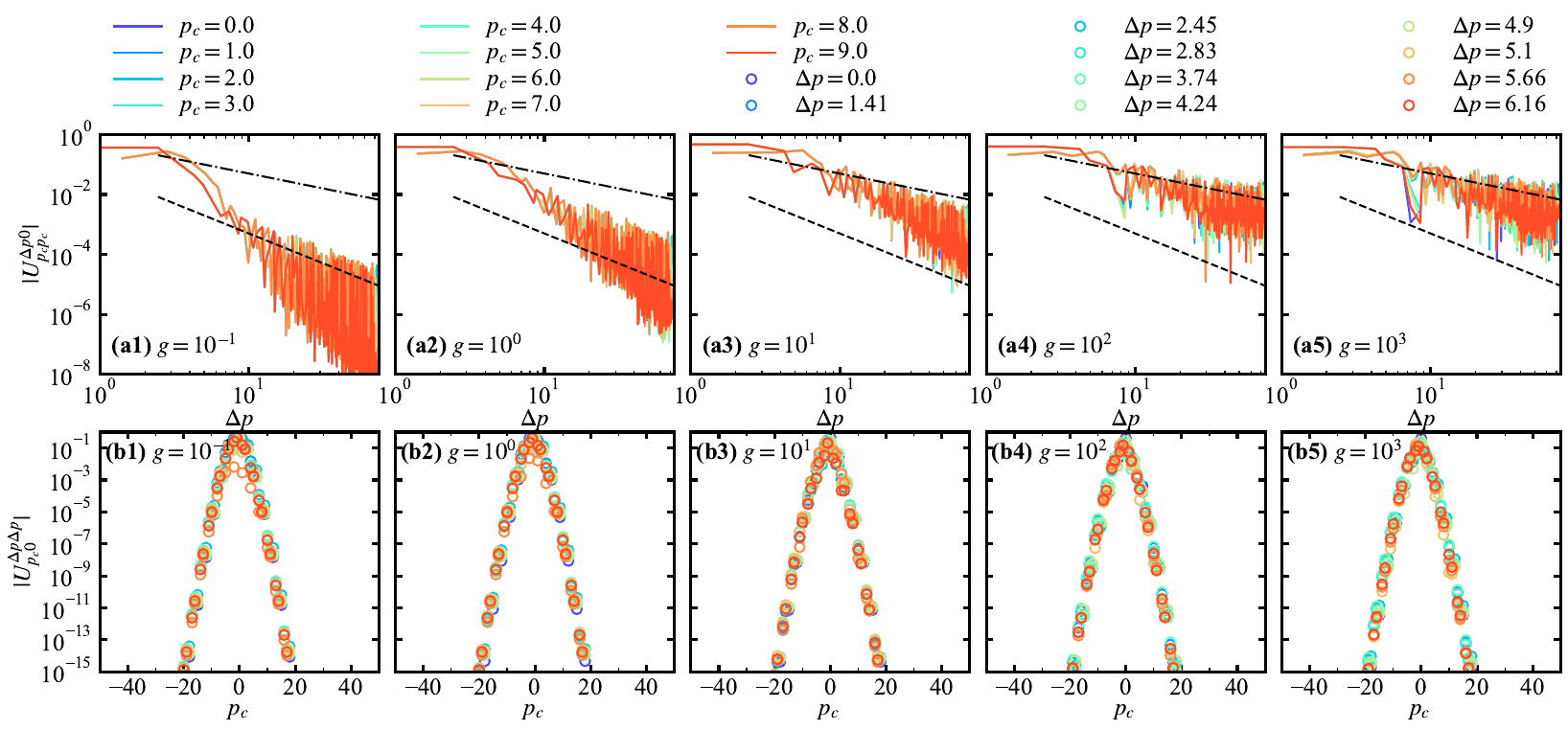}
\caption{
(a1-a5) The hopping amplitude $|U_{p_\mathrm{c}p_\mathrm{c}}^{\Delta p0}|$ as a function of the standard deviation $\Delta p$ for different interactions $g$ in log-log scale. The dashed line and dot-dashed line denote algebraic decay as $\Delta p^{-2}$ and $\Delta p^{-1}$, respectively. (b1-b5) The hopping amplitude $|U_{p_\mathrm{c}0}^{\Delta p\Delta p}|$ as a function of the CM momentum $p_\mathrm{c}$ for different interactions $g$ in a semi-log scale. The particle number, cut-off and kick strength are $N=3$, $M=30$ and $K=1$, respectively.
}
\label{Supp_diff_momentum}
\end{figure}
In the main text, we show a specific case that the hopping amplitude from the BEC state $\lvert0, 0\rangle$, whose kinetic energy is zero. 
In this section, we also consider initial states as highly excited ones $\lvert p_\mathrm{c}\neq 0,  \Delta p=0\rangle$ and $\lvert p_\mathrm{c}= 0,  \Delta p\neq0\rangle$. Figure~\ref{Supp_diff_momentum} shows the hopping amplitude of standard deviation spreading (a1-a5) $|U_{p_\mathrm{c}p_\mathrm{c}}^{\Delta p0}|$ and CM momentum expansion $|U_{p_\mathrm{c}0}^{\Delta p\Delta p}|$ (b1-b5) for different interaction strengths. We find that for various initial momenta, the decay behaviors remain almost identical. With increasing interaction strength, all algebraic tails in $|U_{p_\mathrm{c}p_\mathrm{c}}^{\Delta p0}|$ consistently change from $\Delta p^{-2}$ to $\Delta p^{-1}$. The amplitudes of CM momentum expansion $|U_{p_\mathrm{c}0}^{\Delta p\Delta p}|$ follow the same fast decay. The localization length is almost independent of interactions and initial standard deviation $\Delta p$.

\subsection{The fitting procedure of the generalized fractal dimension}
\begin{figure*}
\centering
\includegraphics[width=0.8\linewidth]{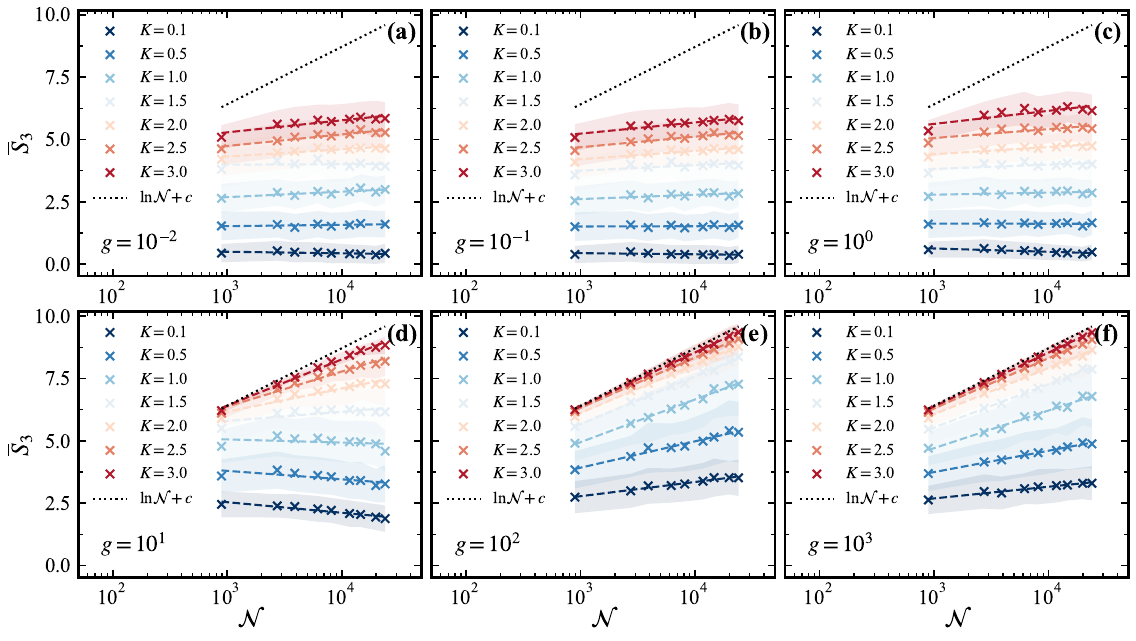}
\caption{
The linear fittings (dashed lines) of $\overline{S}_3$ as a function of dimension $\mathcal{N}$ for different $g$ and $K$. The shaded areas denote the standard deviation obtained by the average over $500$ eigenstates. The number of bosons is fixed as $N=3$.
}
\label{supp_fitting}
\end{figure*}

\begin{figure*}
\centering
\includegraphics[width=0.75\linewidth]{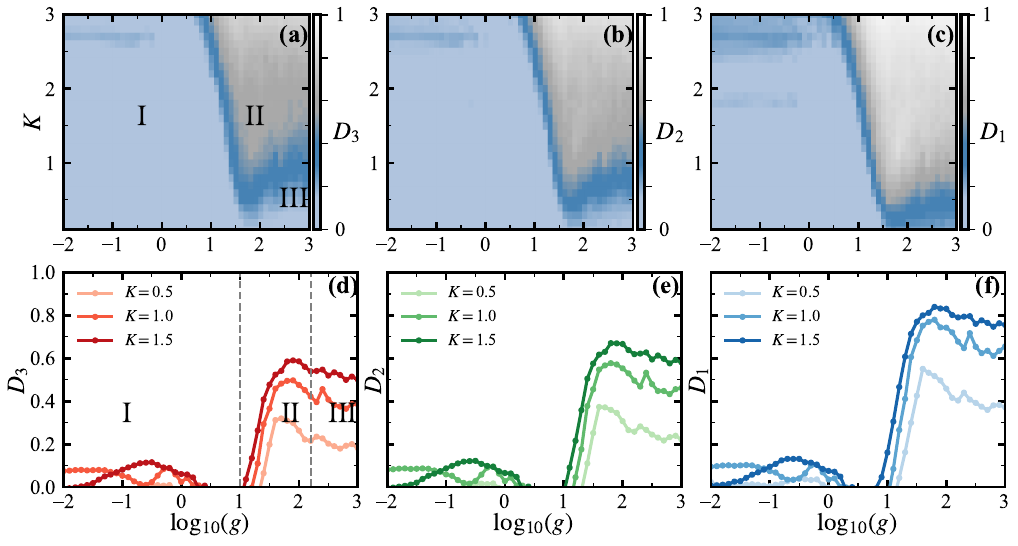}
\caption{
The fractal dimensions (a) $D_3$, (b) $D_2$, and (c) $D_1$ as a function of the interaction strength $g$ and the kick strength $K$, respectively. The number of bosons is $N=3$. The area at large $K$ represent numerical inaccurate regime due to finite truncation. (d-f) Different $D_{\beta}$ as a function of $g$ at different $K$. The gray dashed lines denote the regime boundaries.
}
\label{supp_diff_GFD}
\end{figure*}

\begin{figure*}
\centering
\includegraphics[width=0.75\linewidth]{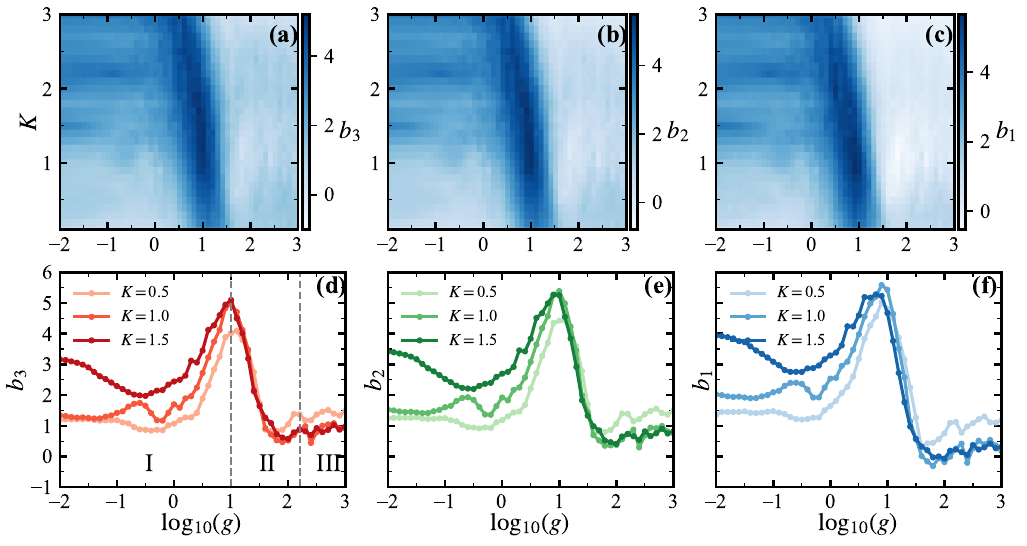}
\caption{
The fitted intercepts (a) $b_3$, (b) $b_2$,  and (c) $b_1$ as a function of the interaction strength $g$ and the kick strength $K$, respectively. The number of bosons is $N=3$. (d-f) Different $b_{\beta}$ as a function of $g$ for different $K$. The gray dashed lines denote the regime boundaries.
}
\label{supp_diff_bq}
\end{figure*}
To identify the possible dominant asymptotic behaviors of the PE $S_{\beta}$ as the dimension $\mathcal{N}$ grows, we utilize a scaling form $\overline{S}_{\beta}=D_{\beta}\mathrm{ln}\mathcal{N}+b_{\beta}$. In particular, the PE at $\beta=1$ is the Shannon entropy $S_1=-\sum_{\alpha}\lvert\psi_{\alpha}\rvert^{2}\mathrm{ln}\lvert\psi_{\alpha}\rvert^{2}$ while $\beta=2$ gives the inverse participation ratio (IPR) with $S_2=-\mathrm{ln}(\mathrm{IPR})$. Note that, a larger kick strength induces a longer localization length $l_{\mathrm{loc}}$ in the truncated momentum space $M$. One needs to push the cutoff towards infinity so that $M\gg l_{\mathrm{loc}}$ to guarantee the convergence, which is quite challenging in ED simulations. This restricts the range of kick and interaction strengths considered, as the condition $g\gtrsim LM^2$ is unphysical for finite $g$.

Figure~\ref{supp_fitting} shows the fitting details of $S_3$ for various interaction strengths and kick strengths. The linear fitting seemingly works very well and the standard deviations are controllable in a small range. The linear fittings of other PE ($S_2, S_1$) are similar. Intuitively, the dimension is always positive, whereas for small $K$ and $g=10^1$ we find anomaly that $D_3, D_2, D_1<0$. This might be attributed to the closeness to the crossover between regimes I and II. Through this approach, we extract the fitted slopes as the fractal dimensions $D_{\beta}$, then obtain their 2D diagram as a function of $g$ and $K$, as shown in Fig.~\ref{supp_diff_GFD}(a-c). One can see that different $D_{\beta}$ gives a roughly similar behavior, in the sense that there are always three regimes separated by different values.  Regime II is remarkable in $D_1$ diagram. If the kick strength $K$ is fixed, we can find the non-monotonic behavior with the increase of $g$ as shown in Fig.~\ref{supp_diff_GFD}(d-f). As $g$ increases, all $D_{\beta=1,2,3}$ jump to a peak value followed by a decrease. For weak interaction regime, say $g\sim 10^{-1}$, the system is close to free bosons, whereas after the peak regime (regime III) the system is fermionization and thus it can be described by free fermions. This implies the origin of the non-monotonic behavior. 

Furthermore, another quantity is the fitted intercept $b_{\beta}$. In the context of many-body localization, the sign of $b_{\beta}$ signifies the boundary between localization and ergodicity~\cite{nicolas2019MBLfractal}. Figure~\ref{supp_diff_bq}(a-c) also shows the fitted intercept $b_{\beta}$ as a function of $g$ and $K$. Interestingly, $b_{\beta}$ exhibits similar crossovers between three regimes as $D_{\beta}$. One can also find the non-monotonic behavior in Figs.~\ref{supp_diff_bq}(d-f), but it is different that $b_{\beta}$ suddenly drops when $D_{\beta}$ increases. The non-monotonic behavior of $D_{\beta}$ and $b_{\beta}$ signifies different characterizations of the MBDL phase. Notably, $b_{\beta}>0$ almost always holds for small $K$, whereas $b_{\beta}<0$ happens for large $K$ and $g$. As we discussed before, the numerical accuracy for large $K$ are limited by the cut-off. At least, there is no ergodicity in the low $K$ regime, say $K<3$. Nevertheless, $b_{\beta}$ is approaching to $0$ at regime II, signifying the scalability nature of the system.

\subsection{The energy level-spacing statistics}
\begin{figure*}
\centering
\includegraphics[width=1\linewidth]{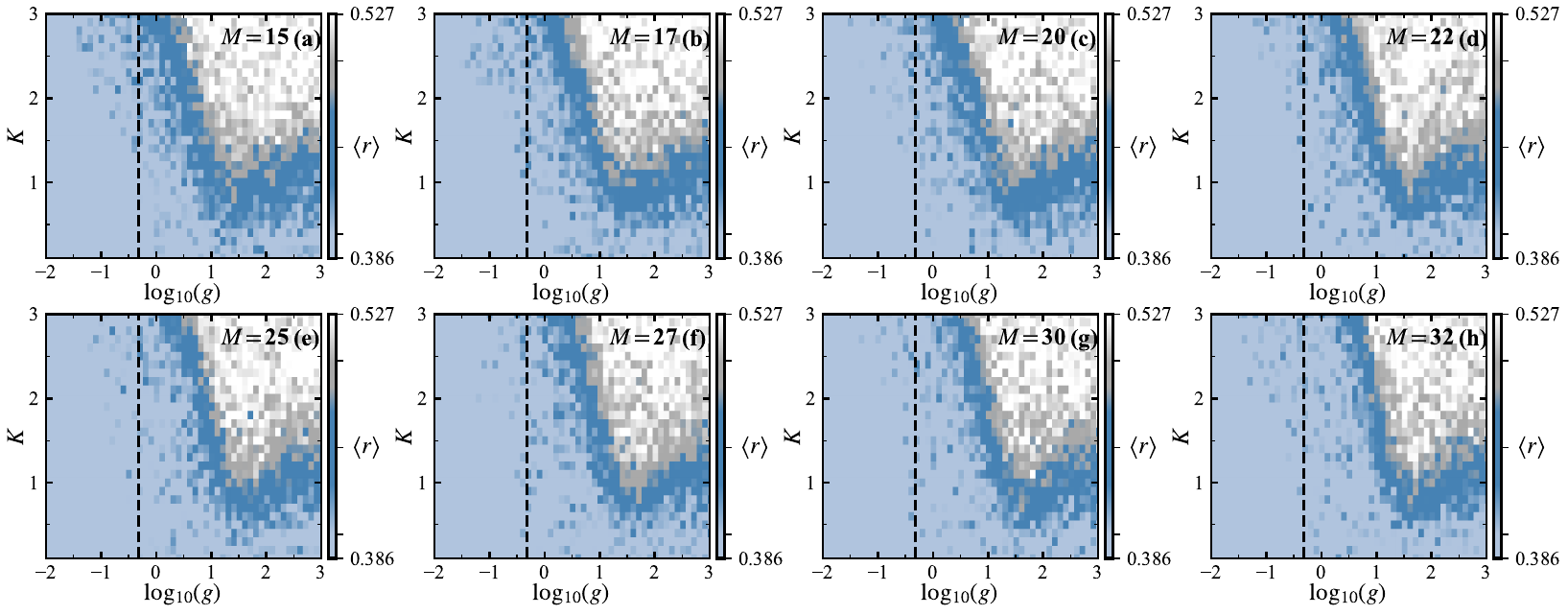}
\caption{
The averaged energy spacing ratio $\langle r\rangle$ as a function of the interaction strength $g$ and the kick strength $K$ for different cut-off $M$. The number of bosons is $N=3$. The area at large $K$ represents numerical inaccurate regime due to finite truncation.
}
\label{supp_diff_r}
\end{figure*}

\begin{figure*}
\centering
\includegraphics[width=1\linewidth]{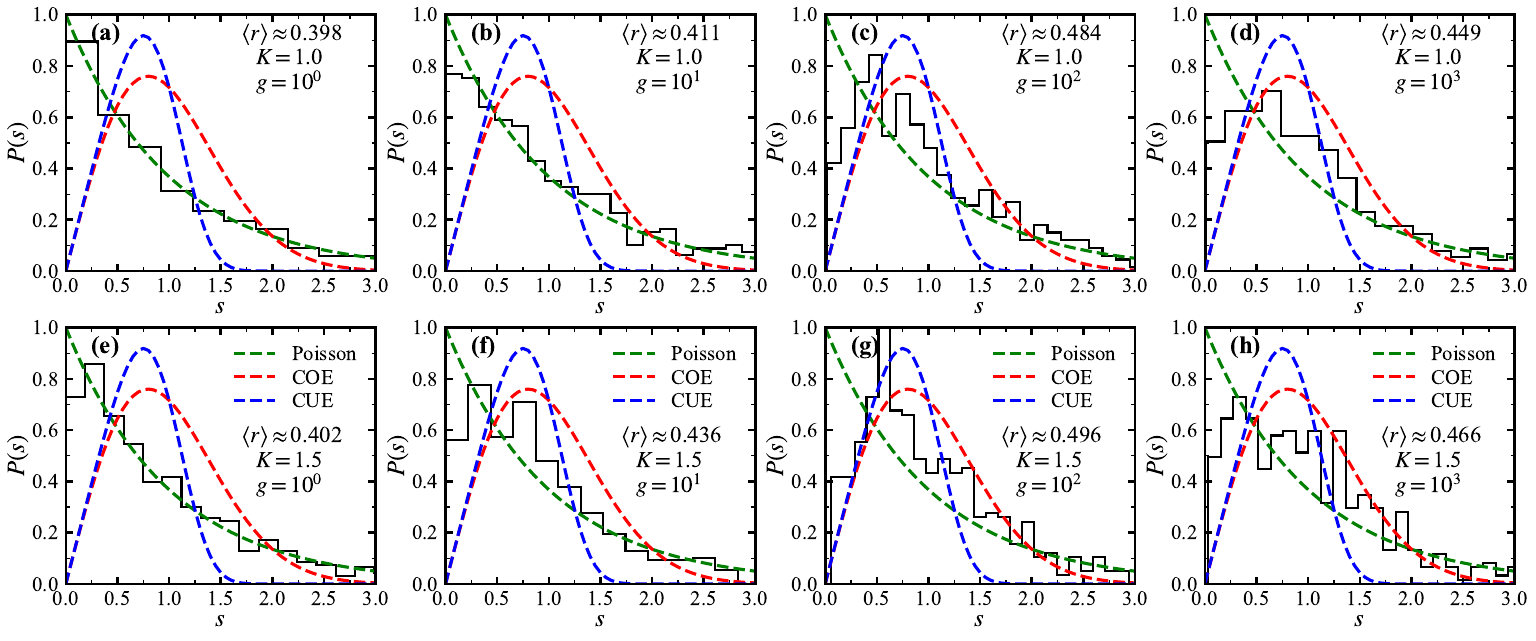}
\caption{
The histograms of the level-spacing distributions $P(s)$ for different kick strength and interactions. The number of bosons and cut-off is $N=3$ and $M=32$, respectively. The green dashed line, the red dashed line and the purple dashed lines denote the Poisson, the COE and the CUE distribution.
}
\label{supp_diff_distribution}
\end{figure*}
Considering the finite cut-off $M$ in momentum space, here we study its effect on the level-spacing statistics in Fig.~\ref{supp_diff_r}. Despite some differences at small $g$ and large $K$, all the $\langle r\rangle$ diagrams exhibit a similar behavior. Note that the weak-interacting bosons are in strong degeneracy, which is solved after $\gamma=1$ (the dashed line). Even for the smallest $M$, there exists a cone (V-shape pattern in $\langle r\rangle$ diagrams) at the TG regime. As $M$ increases, the cone at regime II becomes clearer. Therefore, our results are reasonable even under the finite cut-off and the cone regime is expected to persist as the cut-off is pushed toward infinity.  For the largest $M$, we further pick two different kick strengths to see the details by the level-spacing distribution, which contains more information than $\langle r\rangle$. With the consecutive energy gaps $\delta_{\alpha}=\theta_{\alpha+1}-\theta_{\alpha}$, we count the frequency with which the energy gap occurs in a certain range and plot the histogram. Figure~\ref{supp_diff_distribution} shows the level-spacing distributions $P(s)$ for different $K$ and $g$, where $s=\delta_{\alpha}/\langle \delta_{\alpha}\rangle$. For an integrable system, $P(s)$ follows the Poisson distribution (green). For an ergodic Floquet system with time-reversal symmetry, $P(s)$ follows the circular orthogonal ensemble (COE) type, whereas with broken time-reversal symmetry it follows the circular unitary ensemble (CUE) type~\cite{rigol2014longtimedriven}. According to the random matrix theory, the exact formulas for the above three distributions are as follows~\cite{haake1991quantumchaos}:
\begin{equation}
    P_{\mathrm{Poisson}}(s) = e^{-s}, \quad\quad
    P_{\mathrm{COE}}(s) = \frac{\pi}{2}se^{-\pi s^2/4}, \quad\quad
    P_{\mathrm{CUE}}(s) = \frac{\pi}{2}se^{-\pi s^4/4}.  
\end{equation}
When $g$ is small at regime I, the level-spacing ratio suggest an integrable system. As shown in Fig.~\ref{supp_diff_distribution}(a) and (e), $P(s)$ is well fitted by the Poisson distribution. As $g$ increases, there are clear deviations appearing between $P(s)$ and the Poisson distribution and $P(s)$ show closeness to the COE distribution. However, such deviations become small with further increasing $g$. This non-monotonic behavior aligns with that of $\langle r\rangle$.

\end{document}